\documentclass[11pt]{iopart} % IOP style

\usepackage[dvips]{graphicx}
\usepackage{subfigure}
\usepackage{amsmath}
\usepackage{amssymb}
\usepackage{bm}
\usepackage{color}
\usepackage[colorlinks=true,citecolor=blue,urlcolor=blue,linkcolor=blue,hyperfigures=true]{hyperref}
\usepackage{multirow}

\def\Eq{eq.~}
\def\Eqs{eqs.~}
\def\Fig{figure~}

\def\Ref{ref.~}
\def\Refs{refs.~}
\def\Sec{sec.~}

\def\be{\begin{equation}}
\def\ee{\end{equation}}
\def\bea{\begin{eqnarray}}
\def\eea{\end{eqnarray}}

\def\ie{\textit{i.e.}~}
\def\eg{\textit{e.g.}~}
\def\etal{\textit{et al}~}
\def\diff{\mathrm{d}}
\newcommand{\ket}[1]{\left| #1 \right\rangle}

\begin{document}
%%%%%%%%%%%%%%%%%%%%%%%%%%%%%%%%%%%%%%%%%%%%%%%%%%%%%%%%%%%%%%%%%%%%%%%%%%%%%%%%%%%%%%%%%%%%%%%%%

\title[Correlative methods for quantum tests of the weak equivalence principle]
{Correlative methods for dual-species quantum tests of the weak equivalence principle}

\author{B. Barrett, L. Antoni-Micollier, L. Chichet, B. Battelier, P.-A. Gominet, A. Bertoldi, P. Bouyer}
\address{LP2N, IOGS, CNRS and Universit\'{e} de Bordeaux, rue Fran\c{c}ois Mitterrand, 33400 Talence, France}

\author{A. Landragin}
\address{LNE-SYRTE, Observatoire de Paris, CNRS and UPMC, 61 avenue de l'Observatoire, 75014 Paris, France}

\ead{brynle.barrett@institutoptique.fr}

%================================================================================================
% Abstract
%================================================================================================

\begin{abstract}
Matter-wave interferometers utilizing different isotopes or chemical elements intrinsically have different sensitivities, and the analysis tools available until now are insufficient for accurately estimating the atomic phase difference under many experimental conditions. In this work, we describe and demonstrate two new methods for extracting the differential phase between dual-species atom interferometers for precise tests of the weak equivalence principle. The first method is a generalized Bayesian analysis, which uses knowledge of the system noise to estimate the differential phase based on a statistical model. The second method utilizes a mechanical accelerometer to reconstruct single-sensor interference fringes based on measurements of the vibration-induced phase. An improved ellipse-fitting algorithm is also implemented as a third method for comparison. These analysis tools are investigated using both numerical simulations and experimental data from simultaneous $^{87}$Rb and $^{39}$K interferometers, and both new techniques are shown to produce bias-free estimates of the differential phase. We also report observations of phase correlations between atom interferometers composed of different chemical species. This correlation enables us to reject common-mode vibration noise by a factor of 730, and to make preliminary tests of the weak equivalence principle with a sensitivity of $1.6 \times 10^{-6}$ per measurement with an interrogation time of $T = 10$ ms. We study the level of vibration rejection by varying the temporal overlap between interferometers in a symmetric timing sequence. Finally, we discuss the limitations of the new analysis methods for future applications of differential atom interferometry.
\end{abstract}

\submitto{\NJP}

\maketitle

%================================================================================================
\section{Introduction}
\label{sec:Introduction}

Einstein's equivalence principle (EEP) is a fundamental concept in physics that describes the exact correspondence between the gravitational and inertial mass of any object. It is a central assumption of the theory of General Relativity---which interprets gravity as a geometrical feature of space-time, and predicts identical accelerations for different objects in the same gravitational field. Precise tests of the EEP are of great interest in various fields of physics. For instance, some theories that attempt to unify gravity with the other fundamental forces predict a violation of this principle \cite{Damour2002, Will2006}. The detection of such a violation could aid our understanding of dark energy in cosmology, and advance the search for physics beyond the Standard Model. In contrast, null results are also pivotal for putting bounds on model parameters contained in various extensions to General Relativity \cite{Damour1993a, Damour1993b,Hohensee2013}. The equivalence principle is generally divided into three sub-principles that each must be satisfied for the EEP to hold \cite{Will1993, Altschul2015}: the local Lorentz invariance, the local position invariance and the weak equivalence principle (WEP). In this article, we will focus on the latter.

The WEP---otherwise known as the universality of free fall---states that a charge-free body will undergo an acceleration in a gravitational field that is independent of its internal structure or composition. Tests of the WEP generally involve measuring the relative acceleration between two different test bodies that are in free fall with the same gravitational field. The WEP is characterized by the E\"{o}tv\"{o}s parameter, $\eta$, given by
\be
  \label{eta}
  \eta = 2 \, \frac{a_1 - a_2}{a_1 + a_2} = \frac{\Delta a}{a},
\ee
where $a_1$ and $a_2$ are the accelerations of the two bodies, $\Delta a = a_1 - a_2$ is the relative acceleration, and $a = (a_1 + a_2)/2$ is the average acceleration. The WEP is satisfied if and only if $\Delta a = 0$---implying that $\eta = 0$.

The most precise tests of the WEP have been carried out with lunar laser ranging techniques \cite{Williams2004}, or using a rotating torsion balance \cite{Schlamminger2008,Wagner2012}, which have both measured $\eta$ at the level of a few parts in $10^{13}$. Various Space missions to test the WEP at improved levels ($10^{-15}$ or better) using other classical devices are presently in progress \cite{Nobili2012, Overduin2012, Berge2015}. On a separate frontier, a number of groups have carried out tests between cold atoms \cite{Fray2004, Bonnin2013, Tarallo2014, Schlippert2014, Zhou2015} in an effort to probe the WEP at the quantum level. The majority of these tests have been conducted using matter-wave interferometers which, over the past few decades, have been extensively studied both theoretically and experimentally \cite{Borde1989, Berman1997, Cronin2009, Tino2014}. Atom interferometers have been utilized as ultra-precise inertial sensors to measure, for example, the gravitational acceleration $g$ \cite{Peters1999, Peters2001, Sorrentino2012, Gillot2014}, the gravitational constant $G$ \cite{Fixler2007, Rosi2014, Biedermann2015}, gravity gradients \cite{Sorrentino2012, Snadden1998, McGuirk2002, Wu2009, Sorrentino2014}, gravitational field curvature \cite{Rosi2015}, and rotations \cite{Gustavson1997, Canuel2006, Barrett2014a, Tackmann2014}. A WEP test based on atom interferometry involves measuring the differential phase shift resulting from a relative acceleration between two species with different masses that are in free fall within the same gravitational field. This measurement is based on the same principle as gravity gradiometry, where the quantity of interest is the differential phase between test atoms of the same type but in different spatial locations. The gradient of the gravitational field can be extracted from the differential phase between two sources, while higher derivatives of the field can be accessed if more than two sources are used. This technique was recently demonstrated to measure the curvature of the gravitational field, and has been proposed to detect gravitational waves and to study geophysical effects \cite{Dimopoulos2008, Canuel2014, Barrett2014}. Presently, the state-of-the-art for WEP tests using matter-wave interferometry corresponds to an uncertainty of $3 \times 10^{-8}$ \cite{Zhou2015}. A comparison between the gravitational acceleration measured by atoms and a macroscopic object (\ie a falling corner-cube) have also been carried out, and yield agreement at the level of $\delta \eta \simeq 6.5 \times 10^{-9}$ \cite{Merlet2010}. A handful of ground-based \cite{Dimopoulos2007, Dickerson2013, Sugarbaker2013, Hartwig2015} and micro-gravity-based \cite{Barrett2014, Vogel2006, Varoquaux2009, VanZoest2010, Muntinga2013} cold-atom experiments are currently underway that aim to greatly improve this precision. In addition, there have been a number of proposals for Space-based quantum tests of the WEP \cite{Altschul2015, Sorrentino2011, Tino2013, Schubert2013, Aguilera2014} that target accuracies at the level of $10^{-15}$.

So far, most tests with cold atoms have used two isotopes of the same atomic element, \eg $^{85}$Rb and $^{87}$Rb \cite{Fray2004, Bonnin2013, Zhou2015, Kuhn2014}, or $^{87}$Sr and $^{88}$Sr \cite{Tarallo2014}. Although this class of test bodies has demonstrated a good level of common-mode noise rejection when performing differential phase measurements \cite{Bonnin2013}, it is intrinsically less sensitive to possible violations of the equivalence principle because the two atoms are relatively similar in mass and composition. Thus, it is interesting to perform these tests with two entirely different atomic elements. In this article, we will focus on the case of $^{87}$Rb and $^{39}$K. These atoms exhibit a large difference in their number of nuclei---facilitating a mass ratio of $M_{\rm Rb}/M_{\rm K} \sim 2.2$. Additionally, they have identical hyperfine spin structure, and similar excitation wavelengths (around 780 nm and 767 nm, respectively), which enables the use of the same laser technology and optics for cooling and interferometry. Dual-species interferometers of this type have the added advantage of being highly independent---that is, atomic sample properties such as the size and temperature, or interferometer parameters such as the interrogation time, Raman phase, and detuning, can be controlled independently. In contrast to dual-isotope setups where many of these parameters are coupled, this feature is ideal for studying a variety of systematic effects that will be important for future precision measurements \cite{Kuhn2014}. For a more complete comparison of alkali atoms as candidates for WEP tests, see for example \Ref \cite{Chen2014}.

One complication that arises with non-common elemental species is a difference in the scale factors, $S_j \simeq k_j^{\rm eff} T_j^2$, between the interferometers. When the interrogation times $T_j$ are the same, this difference originates from the effective wave vectors $k_j^{\rm eff}$ of the interferometer beams used for atoms $j = 1$ and 2. Assuming that the WEP is true, the phase shift of the two interferometers due to a common acceleration $a$ is $\Phi_j = S_j a$. Thus, a difference in the scale factors produces a relative phase shift between interferometers for the same acceleration: $\delta \phi_d^{\rm sys} = (S_1 - S_2) a$. For the case of $^{85}$Rb and $^{87}$Rb, the scale factors can be made the same by a suitable choice of Raman laser detuning that guarantees $k_1^{\rm eff} = k_2^{\rm eff}$ \cite{Tino2013}. However, this is not generally possible for different chemical elements, and the systematic phase shift must be addressed in other ways.

Another issue related to having different scale factors regards the rejection of common-mode vibration noise between interferometers. From an analysis of the interferometer transfer functions (see \ref{app:Response} or \Refs \cite{Tino2013, Cheinet2008, Geiger2011phd}, for instance), one can show that perfect common-mode rejection requires four conditions to be satisfied: (i) the interferometers occur simultaneously with $T_1 = T_2$, such that they experience the same vibration noise, (ii) they have identical wave vectors, $k_1^{\rm eff} = k_2^{\rm eff}$, they exhibit (iii) identical effective Rabi frequencies, $\Omega_1^{\rm eff} = \Omega_2^{\rm eff}$, and (iv) identical pulse durations, $\tau_1 = \tau_2$. These conditions imply that if $S_1 \not= S_2$, the interferometers do not respond to common-mode noise with the same phase shift.

The scale factors can be made the same by adjusting the interrogation times of the interferometers such that $T_1 = r T_2$, where $r = \sqrt{k_2^{\rm eff}/k_1^{\rm eff}}$ \cite{Varoquaux2009}. This technique eliminates the systematic phase shift $\delta \phi_d^{\rm sys}$ resulting from a constant acceleration, and improves the rejection of common vibration noise at frequencies $\lesssim 1/T_1$, but it degrades the rejection efficiency at frequencies above $\sim 1/T_1$ (see \ref{app:Response}). However, if the ratio $r$ is very close to unity, as it is for some choices of atoms ($r \simeq 1.009$ for $^{39}$K and $^{87}$Rb), this option represents a good compromise between efficient noise rejection and reducing systematic effects.

In this article, we describe and demonstrate three analysis methods for atom-interferometric WEP tests---including two new techniques that eliminate both aforementioned problems of systematic phase shifts and diminished common-mode rejection between coupled interferometers of different atomic species. The first of these two new methods is a generalized Bayesian analysis of the Lissajous curves formed by plotting the coupled sensor measurements parametrically. The second technique involves restoring the interferometer fringes by correlating with an auxiliary mechanical accelerometer. In this case, the phase shift for each species can be measured directly from the reconstructed fringes regardless of their scale factors or the degree of temporal overlap between the interferometers. Both of these new methods intrinsically account for different scale factors, and return unbiased estimates of the differential phase. Finally, to give a complete picture, we compare these techniques with an improved ellipse-fitting method recently developed by Szpak \etal \cite{Szpak2012}. This numerical procedure yields an estimate of the differential phase shift with reduced bias compared to more commonly implemented algorithms in the presence of significant amounts of uncorrelated noise between sensors.

In this work, we also report correlated phase measurements between simultaneous interferometers of different chemical species ($^{39}$K and $^{87}$Rb). When operated in an environment with significant levels of vibration noise, we demonstrate a common-mode vibration rejection factor of $\gamma \simeq 730$. These results represent a major step toward precise tests of the WEP with elements exhibiting vastly different masses. We also investigate the accuracy of the three aforementioned methods on experimental data obtained from the K-Rb interferometer.

The article is organized as follows. Section \ref{sec:Principle} reviews some theoretical background concerning a WEP test with a dual-species interferometer. In \Sec \ref{sec:Methods}, we briefly describe the three methods of extracting the differential phase. We give a brief description of the experimental setup for the K-Rb interferometer in \Sec \ref{sec:Experiment}. We present our experimental results in \Sec \ref{sec:Results}, and we give a discussion of the advantages and limitations of the new methods in \Sec \ref{sec:Limitations}. Finally, we conclude in \Sec \ref{sec:Conclusion}. A detailed description of the three analysis methods, including extensive numerical tests of the generalized Bayesian estimator, can be found in the Appendices.

%================================================================================================
\section{Testing the WEP with two atomic species}
\label{sec:Principle}

An atom-interferometric test of the WEP involves measuring the relative acceleration between two atoms of different mass. This can be done in one of two ways: (i) the absolute acceleration of each atom, $a_1$ and $a_2$, can be individually measured and subtracted, or (ii) $\Delta a$ can be measured directly from the differential phase, $\phi_d$. In the ideal case, acceleration measurements are performed simultaneously in order to take advantage of correlated noise between sensors---reducing the total uncertainty in $\Delta a$. Since method (ii) involves a direct measurement of $\phi_d$, it intrinsically requires both simultaneity and phase correlation between atomic sensors to reject common-mode noise. Henceforth, two or more atom interferometers that satisfy these conditions are referred to as ``coupled sensors''. Method (i) can be carried out regardless of these two constraints. In this section, we outline some theoretical background related to a WEP test with method (ii).

Generally, the output from two coupled atomic sensors is described by the following sinusoids
\begin{subequations}
\label{yj(a)}
\begin{align}
  y_1(a) & = A_1 \cos(S_1 a + \phi_1) + B_1, \\
  y_2(a) & = A_2 \cos(S_2 a + \phi_2) + B_2,
\end{align}
\end{subequations}
where $A_j$ and $B_j$ are, respectively, the amplitude and offset of the interferometer fringes associated with sensor $j$ ($j = 1,2$). In principle, these two parameters can be measured and \Eqs \eqref{yj(a)} can be recast in the normalized form $n_j = (y_j - B_j)/A_j$:
\begin{subequations}
\label{nj(a)}
\begin{align}
  n_1(a) & = \cos(S_1 a + \phi_1), \\
  n_2(a) & = \cos(S_2 a + \phi_2).
\end{align}
\end{subequations}
Here, $a$ is an acceleration common to both atoms, $S_j$ is the scale factor for interferometer $j$, and $\phi_j$ is a phase shift. The scale factors can be computed exactly from the integral of the response function, $f_j(t)$, given by \Eq \eqref{fj(t)}:
\be
  S_j = k_j^{\rm eff} \int f_j(t) \diff t = k_j^{\rm eff} (T_j + 2\tau_j) \left( T_j + \frac{4\tau_j}{\pi} \right),
\ee
where $k_j^{\rm eff}$ is the effective wave-vector for the counter-propagating interferometer beams, $T_j$ is the interrogation time, and $\tau_j$ is the $\pi/2$ Raman pulse duration. A detailed explanation of the response function and its role in WEP tests is outlined in \ref{app:Response}. For large interrogation times, $T_j \gg \tau_j$, the scale factors reduce to the well-known relation $S_j \simeq k_j^{\rm eff} T_j^2$.

The phases $\phi_j$ each have three contributions, one from the interferometer laser phase $\phi_j^{\rm laser}$, one from parasitic systematic effects $\phi_j^{\rm sys}$, and one from the WEP signal $\phi_j^{\rm WEP}$
\be
  \phi_j = \phi_j^{\rm laser} + \phi_j^{\rm sys} + \phi_j^{\rm WEP}.
\ee
In an experiment, the laser phase is a control parameter which can be set to zero, and systematic effects are independently nullified as much as possible. The shift due to a WEP violation can be defined as $\phi_j^{\rm WEP} = S_j (a_j - a)$, which is expected to be very close to zero. In the ideal case, the total interferometer phase $\Phi_j$ contains only the shift due to the mean acceleration, $S_j a$, and a WEP violation. It then follows that
\be
  \label{TotalPhij}
  \Phi_j = S_j a + \phi_j = S_j a_j.
\ee
In this case, the total phase of each interferometer can be related to the E\"{o}tv\"{o}s parameter in the following way
\be
  \label{eta2}
  \eta = \frac{\Phi_1/S_1 - \Phi_2/S_2}{a} = \frac{\Delta a}{a}.
\ee
In principle, the sensitivity in this type of WEP test increases as the square of the interrogation time, $T \sim T_1 \sim T_2$, due to the scale factors, $S_j$, that appear inversely in \Eq \eqref{eta2}.

The general form of \Eqs \eqref{nj(a)} describes a Lissajous curve. For the purposes of this analysis, it is useful to redefine the phases in \Eqs \eqref{nj(a)} to reduce the number of free parameters. Choosing sensor 2 as a reference to rescale the phase of sensor 1, we define a common phase $\phi_c$ that satisfies
\begin{subequations}
\label{rescale}
\begin{align}
  \phi_c & \equiv S_2 a + \phi_2, \\
  \kappa \phi_c + \phi_d & \equiv S_1 a + \phi_1.
\end{align}
\end{subequations}
Here, we introduce two new parameters---the scale factor ratio $\kappa$ and the differential phase $\phi_d$---which are constrained from \Eq \eqref{rescale} to be
\be
  \label{phid&alpha}
  \phi_d = \phi_1 - \kappa \phi_2, \;\;\;\;\;\;\; \kappa = \frac{S_1}{S_2}.
\ee
The sensor outputs are now recast with $\phi_c$ as the primarily parameter
\begin{subequations}
\label{nj(phic)}
\begin{align}
  n_1(\phi_c) & = \cos(\kappa \phi_c + \phi_d), \\
  n_2(\phi_c) & = \cos(\phi_c).
\end{align}
\end{subequations}
Comparing \Eqs \eqref{TotalPhij}, \eqref{eta2} and \eqref{phid&alpha}, it follows that the E\"{o}tv\"{o}s parameter is directly proportional to the differential phase:
\be
  \label{eta3}
  \eta = \frac{\phi_d}{S_1 a}.
\ee

%================================================================================================
\section{Correlative methods of differential phase extraction}
\label{sec:Methods}

In this section, we review three different methods to measure the differential phase from experimental data: ellipse fitting, Bayesian analysis and fringe reconstruction from mirror acceleration measurements.

%------------------------------------------------------------------------------------------------
\subsection{Improved ellipse fitting}

The ellipse fitting technique was first applied to atom interferometry in \Ref \cite{Foster2002} for situations in which the phase common to two coupled atomic sensors is sufficiently scrambled to impede individual fringe observation. In this case, when the measurements from each sensor are plotted parametrically, one obtains an ellipse that is free from common phase noise. Using a least-squares ellipse fitting algorithm, the differential phase $\phi_d$ can be extracted. Multiple groups have demonstrated the utility of ellipse fitting for measurements of gravity gradients \cite{Wu2009, Sorrentino2014} and the gravitational constant $G$ \cite{Fixler2007, Rosi2014, Lamporesi2008}. However, this technique suffers from a number of drawbacks. First, it is valid only for coupled sensors with the same scale factor ($\kappa = 1$). Second, in the presence of moderate amounts of noise in the fringe offsets or amplitudes [the parameters $A_j$ and $B_j$ in \Eqs \eqref{yj(a)}], or in the differential phase, the ellipse fit returns a biased estimate of $\phi_d$ \footnote{Rosi \etal \cite{Rosi2015} demonstrated that the bias in the estimate of $\phi_d$ can be eliminated under certain conditions when fitting an ellipse in three dimensions from the output of three simultaneous interferometers.}.

Recently, Szpak \etal \cite{Szpak2012} developed an algorithm based on the optimization of the approximate maximum likelihood distance which seeks a balance between costly geometric methods and stable algebraic techniques. This algorithm---termed the ``fast guaranteed ellipse fitting'' (FGEF) method---exhibits a smaller bias in the differential phase estimate over a relatively large phase range (centered on $\pi/2$) compared to the more commonly used ``direct ellipse fit'' (DEF) technique \cite{Fitzgibbon1999}. Additionally, \Ref \cite{Szpak2014} includes error estimations for the geometrically meaningful ellipse parameters (center coordinates, axes and orientation). We have extended their work to include an estimate of the statistical uncertainty in the differential phase, $\delta \phi_d$. We provide a more detailed comparison between DEF and FGEF methods of ellipse fitting in \ref{app:Ellipse}.

%------------------------------------------------------------------------------------------------
\subsection{Generalized Bayesian analysis}
\label{sec:Bayesian}

Heuristic approaches to estimating the differential phase, such as ellipse-fitting methods, do not have knowledge of the noise present in experimental data, nor of how various types of noise can affect the outcome of measurements. Bayesian analysis offers an efficient alternative to the problem by constraining the estimate based on a statistical model that describes the distribution of data that results from different noise sources \cite{Gelman2003}. Bayesian phase estimation was studied in the context of atom interferometry in \Ref \cite{Stockton2007} for two sensors containing the same scale factor ($\kappa = 1$). In that work, a detailed study of each possible noise source (amplitude, offset and differential phase) is presented. Reference \cite{Chen2014} also used Bayesian analysis to estimate the differential phase from a hypothetical system with $\kappa < 1$. There, however, only noise in the differential phase is considered, and the range of common phase was constrained to $\phi_c \in [0,\pi]$. To the best of our knowledge, no complete Bayesian estimator exists that (i) is valid for any scale factor ratio, (ii) accounts for noise in all relevant system parameters, and (iii) allows $\phi_c$ to vary over a broad range. Furthermore, this type of analysis has not yet been demonstrated on experimental data from dual-species interferometers.

In this work, we have developed a generalized Bayesian estimator for $\phi_d$---based on the approach of \Ref \cite{Stockton2007}---that satisfies all three of the requirements mentioned above. We demonstrate this technique by measuring $\phi_d$ from both simulated data (see \ref{app:Bayesian}) and experimental data from our K-Rb interferometer (see \Sec \ref{sec:Results}). The advantage of using this estimation technique is that the uncertainty in $\phi_d$ converges much faster than other methods (\ie it scales as $\sim 1/\sqrt{N}$, where $N$ is the number of measurements), so fewer data are required to reach a given level of sensitivity. Furthermore, since $\kappa$ is built directly into the Bayesian estimate of $\phi_d$, it is free from the aforementioned systematic phase shift $\delta \phi_d^{\rm sys}$ arising between interferometers with different scale factors. However, some of the drawbacks of the Bayesian analysis are that it requires a priori knowledge of the noise in the system, and it is computationally costly due to the large number of integrals that must be evaluated.

%--------------------------------------------
\begin{figure}[!tb]
  \centering
  \includegraphics[width=0.96\textwidth]{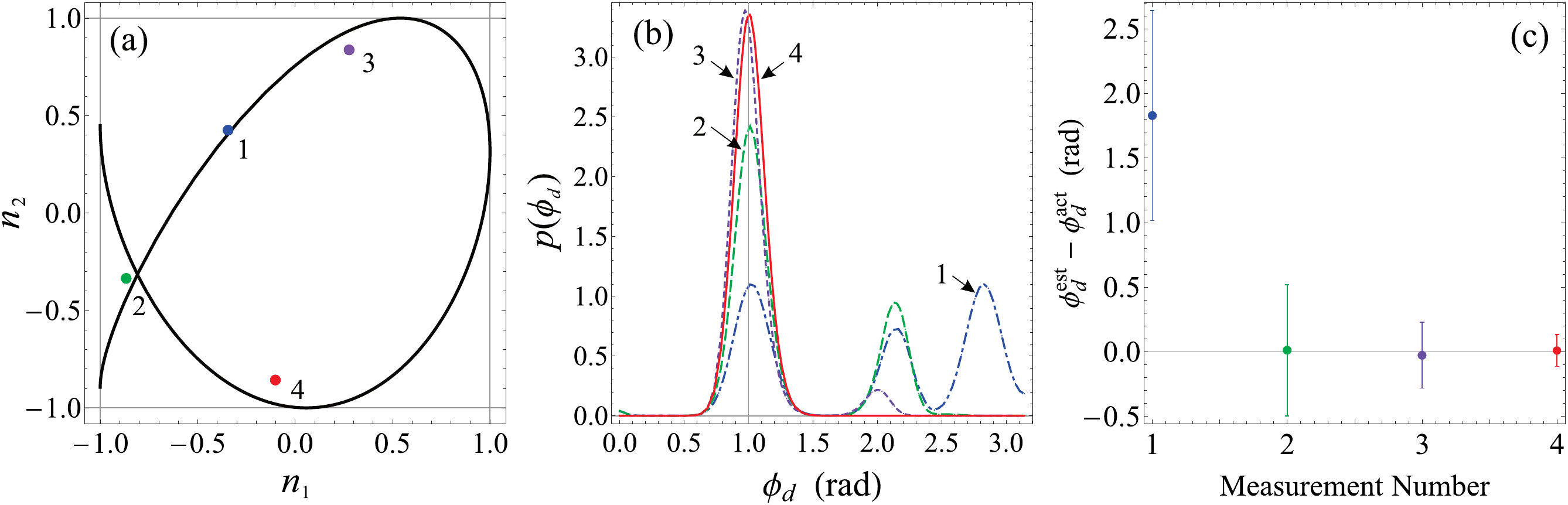}
  \caption{The process of Bayesian estimation of the differential phase $\phi_d$ from synthetic data following a Lissajous curve. (a) Randomly chosen points (labelled $1 - 4$) following \Eqs \eqref{nj(phic)} with Gaussian noise added to $n_1$ and $n_2$. For the actual Lissajous curve (shown as the black curve), we chose $\kappa = 0.8$ and $\phi_d^{\rm act} = 1$ rad for illustrative purposes. (b) The prior probability distribution computed from Bayes' algorithm after each measurement. The vertical solid line indicates the differential phase used in the simulation. (c) Error in the Bayesian estimate after successive measurements. Points represent the difference between $\phi_d^{\rm est}$ and $\phi_d^{\rm act}$ (\ie the systematic error), where $\phi_d^{\rm est}$ is the Bayesian estimate based on the maximum likelihood value from the corresponding prior distribution. The error bars indicate the statistical uncertainty, which are computed from the standard deviation of the prior distributions shown in (b).}
  \label{fig:BayesianEstimationExample}
\end{figure}
%--------------------------------------------

Figure \ref{fig:BayesianEstimationExample} illustrates the basic Bayesian estimation procedure. Here, we simulate data that follow the Lissajous equations \eqref{nj(phic)} with added Gaussian noise in the sensor offsets. After each successive measurement from the system, the width of the new ``prior'' probability distribution decreases and additional peaks are suppressed---facilitating an improvement in the estimate of $\phi_d$. This is how the Bayesian method builds in information from previous measurements. It is clear from \Fig \ref{fig:BayesianEstimationExample}(c) that after only a few iterations, both the statistical and systematic error in $\phi_d$ have decreased dramatically. A detailed description of the generalized Bayesian analysis can be found in \ref{app:Bayesian}.

%------------------------------------------------------------------------------------------------
\subsection{Fringe reconstruction by accelerometer correlation -- The differential FRAC method}
\label{sec:FRAC}

Differential atom interferometry is often utilized under conditions where each sensor is overwhelmed by external phase noise that is common to both sensors. Typically, one is concerned with only the differential phase and not the common phase $\phi_c$, which is treated as an arbitrary parameter. Both the ellipse-fitting and Bayesian estimation methods for extracting $\phi_d$ take this approach. An alternative technique involves measuring the common phase and correcting for it. For the case of parasitic mirror vibrations, single-sensor interference fringes that are otherwise smeared by phase noise can be restored based on measurements from seismometers \cite{LeGouet2008, Merlet2009, Farah2014} or mechanical accelerometers \cite{Barrett2014, Geiger2011, Lautier2014}. Henceforth, we refer to this as the fringe reconstruction by accelerometer correlation (FRAC) method. In this work, we demonstrate how the FRAC method can be applied to two quasi-simultaneous interferometers of different atomic species to measure the relative phase shift between them. This technique to extract $\phi_d$ is referred to as the \emph{differential} FRAC method throughout the article to differentiate between the (standard) FRAC method, which is generally employed to measure the absolute phase shift of a single atom interferometer.

%--------------------------------------------
\begin{figure}[!t]
  \centering
  \includegraphics[width=0.60\textwidth]{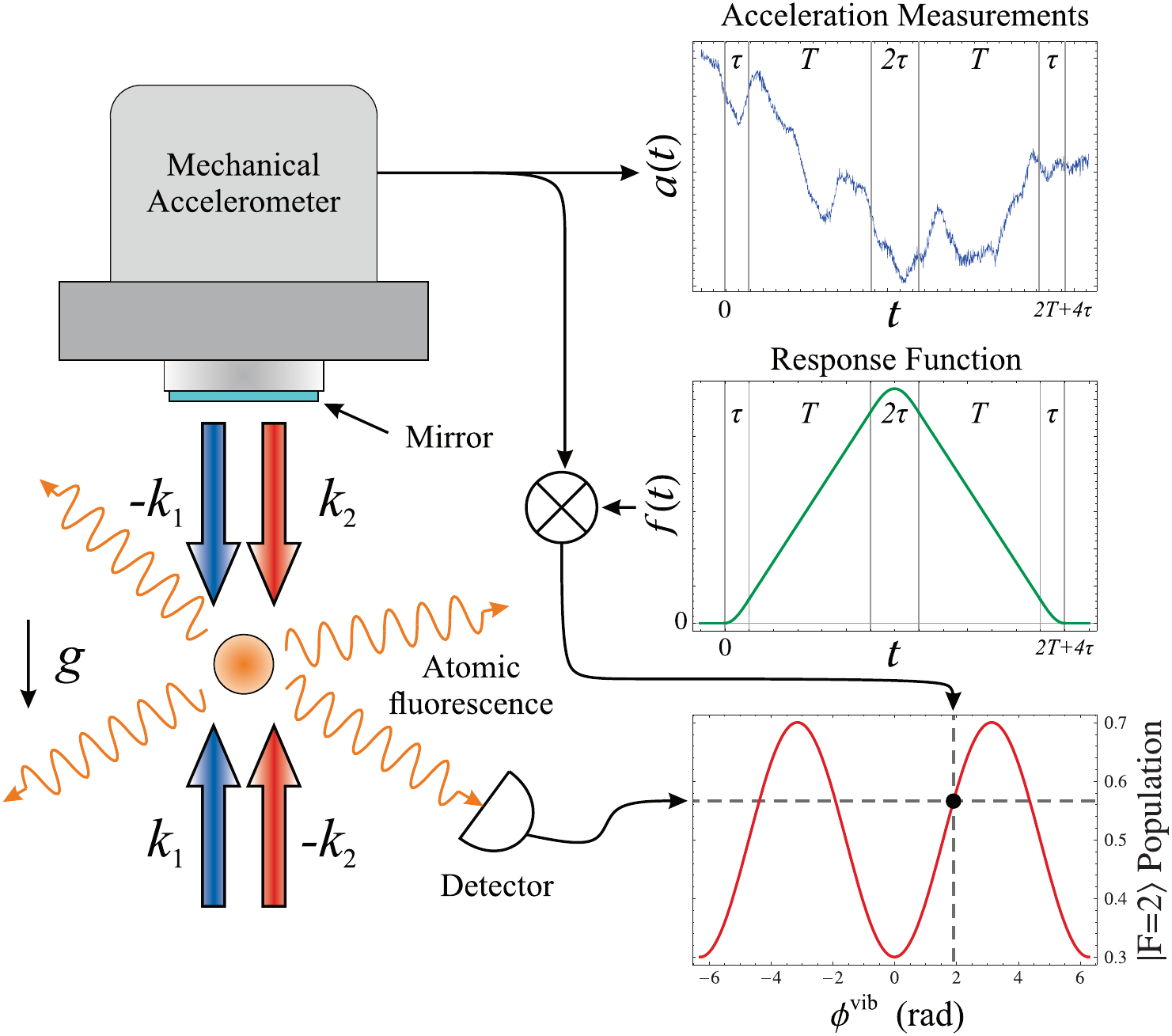}
  \caption{Schematic of the FRAC method for a single interferometer. A mechanical accelerometer is mounted to the back of the retro-reflection mirror for the Raman beam (with wave vectors $k_1$ and $k_2$). Acceleration measurements during the interferometer sequence are weighted by the response function $f(t)$ [see \Eq \eqref{fj(t)}] and integrated to obtain the phase estimate $\phi^{\rm vib}$. Correlating this phase with the interferometer signal during the same time interval reproduces the interference fringe.}
  \label{fig:AccelerometerCorrelation}
\end{figure}
%--------------------------------------------

\begin{figure}[!t]
  \centering
  \subfigure{\includegraphics[width=0.90\textwidth]{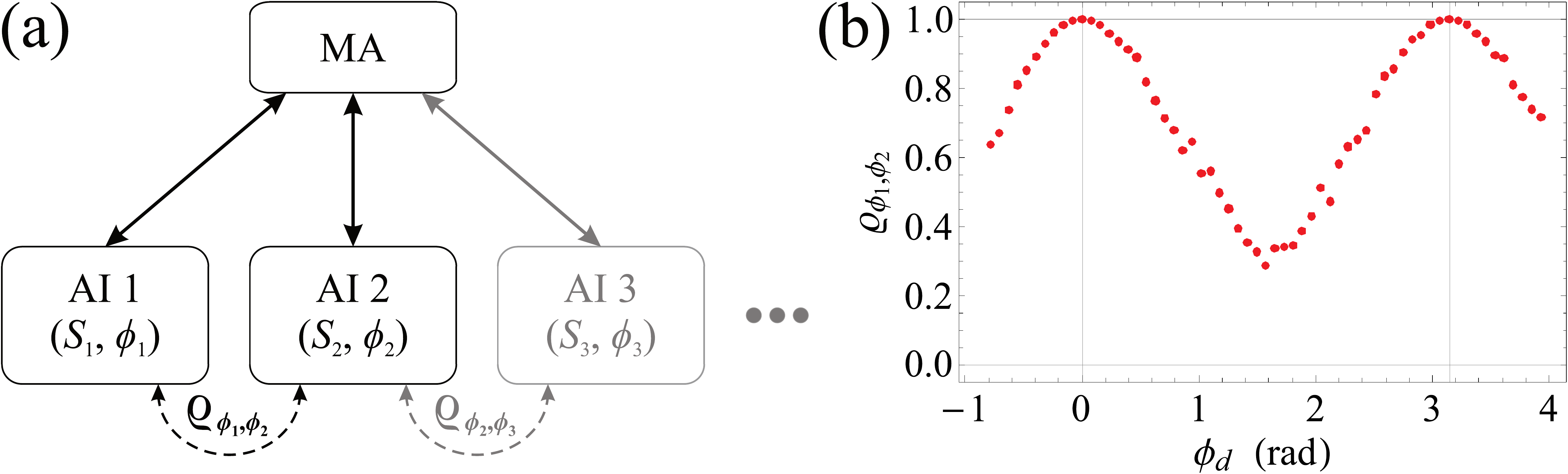}}
  \caption{(a) Schematic illustrating the source of correlation in the differential FRAC method. The signal from each atom interferometer (AI), exhibiting a known scale factor $S_j$ and unknown phase shift $\phi_j$, is directly correlated (indicated by the solid lines) with a common vibration phase measured by a mechanical accelerometer (MA). Since the resulting fringes are derived from a common source, measurements of each $\phi_j$ are highly correlated, as indicated by the dotted lines and characterized by the positive coefficient $\varrho_{\phi_i, \phi_j}$. This technique can be extended for multiple coupled interferometers (shown in gray), although we focus on the case of only two. (b) Correlation coefficient $\varrho_{\phi_1, \phi_2}$ as a function of $\phi_d$ estimated from a large sample of simulated fringes. Here, we assumed $\kappa = 1$ and we added non-common Gaussian noise to the phase of the reconstructed fringes with a standard deviation of $\sigma_{\phi_d} = 0.1$ rad.}
  \label{fig:FRACCorrelation}
\end{figure}

Figure \ref{fig:AccelerometerCorrelation} illustrates the basic schematic of the FRAC method for a single interferometer. A mechanical accelerometer is secured to the back of the reference mirror used to retro-reflect interferometry light, and the time-dependent mirror acceleration, $a^{\rm vib}(t)$, is recorded during the interferometer sequence. These acceleration measurements are first weighted by the response function of the $j^{\rm th}$ interferometer, $f_j(t)$, and are then integrated to find the vibration-induced phase given by\footnote{The underlying assumption of this technique is that motion of the reference mirror at frequencies within the response bandwidth of the mechanical accelerometer are responsible for phase changes of the atom interferometer. Although the corresponding acceleration signal is indistinguishable from fluctuations in local gravity (as a consequence of the equivalence principle), we can be confident in our assumption since typical variations in gravity occur on timescales much larger than $2T$.}
\be
  \label{phijvib}
  \phi_j^{\rm vib} = k_j^{\rm eff} \int f_j(t) a^{\rm vib}(t) \diff t.
\ee
For each repetition of the experiment, this random phase is computed and correlated with the interferometer signal. This process allows one to reconstruct the interference fringes point-by-point. Depending on the level of vibrations and the interferometer sensitivity, the range of vibration-induced phases can span multiple fringes---enabling the single-sensor phase shift $\phi_j$ to be measured using, for instance, a sinusoidal least-squares fit to the data. It is straightforward to extend this algorithm for two or more interferometers, which do not need to be overlapped in time. In this case, the only additional requirement is that the time-series of mirror acceleration measurements span the interrogation times for all interferometers. For two coupled sensors, the differential phase is easily computed from the individual sensor phase shifts via $\phi_d = \phi_1 - \kappa \phi_2$. The statistical error in this quantity is governed by
\be
  \label{deltaphid}
  (\delta \phi_d)^2 = (\delta \phi_1)^2 + (\kappa \delta \phi_2)^2 - 2 \kappa \varrho_{\phi_1, \phi_2} (\delta \phi_1)(\delta \phi_2),
\ee
where the $\delta \phi_j$ represent the statistical uncertainties in the $\phi_j$ obtained from fits to the two fringes, and $\varrho_{\phi_1, \phi_2}$ is the correlation coefficient for the measurements of $\phi_1$ and $\phi_2$. In the limit of perfect correlation ($\varrho_{\phi_1, \phi_2} = 1$), the uncertainty in the differential phase reduces to $\delta \phi_d = |\delta \phi_1 - \kappa \delta \phi_2|$. Figure \ref{fig:FRACCorrelation}(a) illustrates how the coupled-interferometer correlation is utilized by the differential FRAC method. Since the fringes for each interferometer are recovered using measurements from the same classical device, the phase noise present on each fringe is highly correlated. This induces a correlation between the measurements of $\phi_1$ and $\phi_2$ extracted from the fits, as characterized by $\varrho_{\phi_1, \phi_2}$. The key to the differential FRAC method is maximizing this correlation to reduce the uncertainty in $\phi_d$.

%--------------------------------------------
%--------------------------------------------

The correlation coefficient for a given set of reconstructed fringes can be estimated numerically from a large sample of simulated data. We find that it is sensitive to experimental parameters such as the level of uncorrelated noise on each sensor, the scale factor ratio and the differential phase. For instance, \Fig \ref{fig:FRACCorrelation}(b) shows the dependance of $\varrho_{\phi_1, \phi_2}$ on $\phi_d$ for synthetic fringes that contain non-common phase noise with a standard deviation of $0.1$ rad. The correlation coefficient yields a maximum when the interferometers are perfectly in-phase or $\pi$ radians out-of-phase. This is an ideal feature for WEP tests, since the maximum sensitivity occurs exactly at the expected signal of $\phi_d = 0$. This implies that, unlike ellipse-fitting methods where the sensitivity is optimized at $\phi_d = \pi/2$, one does not need to engineer an additional phase shift between the atoms to optimize the sensitivity and reduce systematic bias. Furthermore, a recent study of a gradiometer configuration (\ie $\kappa = 1$) has shown that the differential FRAC method can reach sensitivities close to the quantum-projection-noise limit when modest levels of uncorrelated phase noise are present \cite{PereiraDosSantos2015}.

A number of ideal features make this technique interesting for both absolute and differential atom interferometry experiments.
\begin{itemize}
  \item[1)] The differential FRAC estimate of $\phi_d$ is precise and unbiased over the full phase range $\phi_d \in [0,\pi]$, since it relies on least-squares fits to individual fringes.
  \item[2)] It is simple, fast, and computationally low in cost---allowing the interferometer phase to be corrected in real-time \cite{Lautier2014}, or by post-processing the data \cite{LeGouet2008, Merlet2009, Geiger2011}.
  \item[3)] Unlike the Bayesian analysis, the FRAC method does not require any a priori information about the interferometer offsets, contrasts, and noise parameters---which can be challenging to measure accurately without phase stability \cite{Menoret2012}.
  \item[4)] Systematic phase shifts in $\phi_d$ due to non-identical pulse durations $\tau_j$ and Rabi frequencies $\Omega_j^{\rm eff}$ \cite{Bonnin2013} are accounted for in the estimates of $\phi_j^{\rm vib}$ for each interferometer. Such systematics will be important to consider in future long-baseline differential interferometry experiments \cite{Altschul2015, Dickerson2013, Sugarbaker2013, Hartwig2015, Sorrentino2011, Tino2013, Schubert2013, Aguilera2014}.
  \item[5)] The relative timing between coupled interferometers can be freely chosen---they need not be overlapped. This is a unique feature to dual-species interferometers that do not share the same Raman beams. Unlike the ellipse-fitting and Bayesian techniques, the FRAC method allows one to extract absolute phase information from \emph{each} sensor. Varying the temporal overlap between interferometers can be useful for studying a variety of effects, such as the level of correlation between sensors, or systematics related to the interaction between atoms \cite{Kuhn2014}.
  \item[6)] Single-sensor fringes can be accurately measured in ``noisy'' environments, which is ideal for mobile sensors such as atomic gravimeters \cite{Gillot2014, Barrett2014, Farah2014, Geiger2011}.
\end{itemize}
Although the standard FRAC method is conceptually simple to implement, the drawback is that it is sensitive to errors in the measurements of vibrations. Such errors include the quality of coupling between the mirror and the mechanical device, electronic noise in the signal acquisition, the level of self-noise of the device, drifts in the offset or sensitivity factor, and non-linearities in both the amplitude and frequency response. The natural low-pass filtering feature of atom interferometers can alleviate some of these effects---particularly those that dominate at frequencies beyond the cut-off frequency $1/2T_j$. Alternatively, the mechanical accelerometer can be used as a course measurement of the phase to identify the correct interferometer fringe. Then one can use the atom interferometer output to refine the phase measurement \cite{Geiger2011, Lautier2014} by inverting the sinusoidal relation \eqref{yj(a)}. On the other hand, measurements of $\phi_d$ using the \emph{differential} FRAC method are much less sensitive to many of these noise sources since they are common to two simultaneous interferometers. We discuss the limitations of this method in more detail in \Sec \ref{sec:Limitations}.

%================================================================================================
\section{Description of the ICE experiment}
\label{sec:Experiment}

ICE (Interf\'{e}rom\'{e}trie Coh\'{e}rente pour l'Espace) is an experiment that aims to measure $\eta$ using a dual-species interferometer of $^{87}$Rb and $^{39}$K. It is designed to be transportable and to operate in the micro-gravity environment provided by the Novespace Zero-$g$ plane \cite{Barrett2014, Varoquaux2009, Geiger2011, Nyman2006}. In this section, we give a brief description of the experimental setup.

A detailed description of the telecom-frequency fiber-based laser system used on ICE can be found in \Refs \cite{Barrett2014, Menoret2011}. For each atomic species, we utilize a master-slave architecture, where the master laser diode is locked to either a saturated absorption peak (in the case of rubidium), or to a frequency comb (in the case of potassium). The slave lasers are frequency-locked to their corresponding master through an optical beat-note in the 1550 nm telecom band. After second harmonic generation to 780 nm for $^{87}$Rb and 767 nm for $^{39}$K, the frequency of each slave laser can be precisely adjusted over $\sim 1.3$ GHz within $\sim 2$ ms of settling time. Approximately 1.5 W of total light is available in each slave beam before entering a free-space optical bench. This module is composed of a series of shutters and acousto-optic modulators (AOMs) that are used to split, pulse and frequency shift the light appropriately for cooling, state preparation, interferometry and detection. Finally, the 780 and 767 nm light is coupled into a series of single-mode, polarization-maintaining fibers and sent to the vacuum chamber. The two frequencies required for cooling and repumping, as well as driving Raman transitions in $^{87}$Rb, are generated via a broadband fiber-based electro-optic modulator operating near 6.8 GHz. Similarly, an AOM operating in dual-pass configuration at $\sim 230$ MHz is used to generate these frequencies for $^{39}$K.

The sensor head is composed of a non-magnetic titanium vacuum chamber surrounded by a $\mu$-metal shield. The chamber resides within three nested Helmholtz coils used to compensate residual magnetic fields and to generate a bias along the vertical axis. A custom 2-to-6 way fiber splitter is used to combine the 780 and 767 nm light intended for laser cooling without significant power loss via a polarizing cube and a dichroic wave plate. The splitter subsequently divides the light equally into six beams that are re-coupled into independent fibers used for the dual-species vapor-loaded magneto-optical trap (MOT). In a similar way, light for both detection and interferometry is overlapped in a free-space 2-to-1 way fiber combiner for 780 and 767 nm. The $\sim 2$ cm diameter beams output from the combiner have the same linear polarization, and are aligned along the vertical direction through the atoms. A quarter-wave plate (fabricated for the intermediate wavelength of 773 nm and mounted in front of the retro-reflection mirror) rotates the polarization of the Raman beams by $90^{\circ}$ such that the counter-propagating fields have lin$\perp$lin polarization.

%--------------------------------------------
\begin{figure}[!t]
  \centering
  \includegraphics[width=0.60\textwidth]{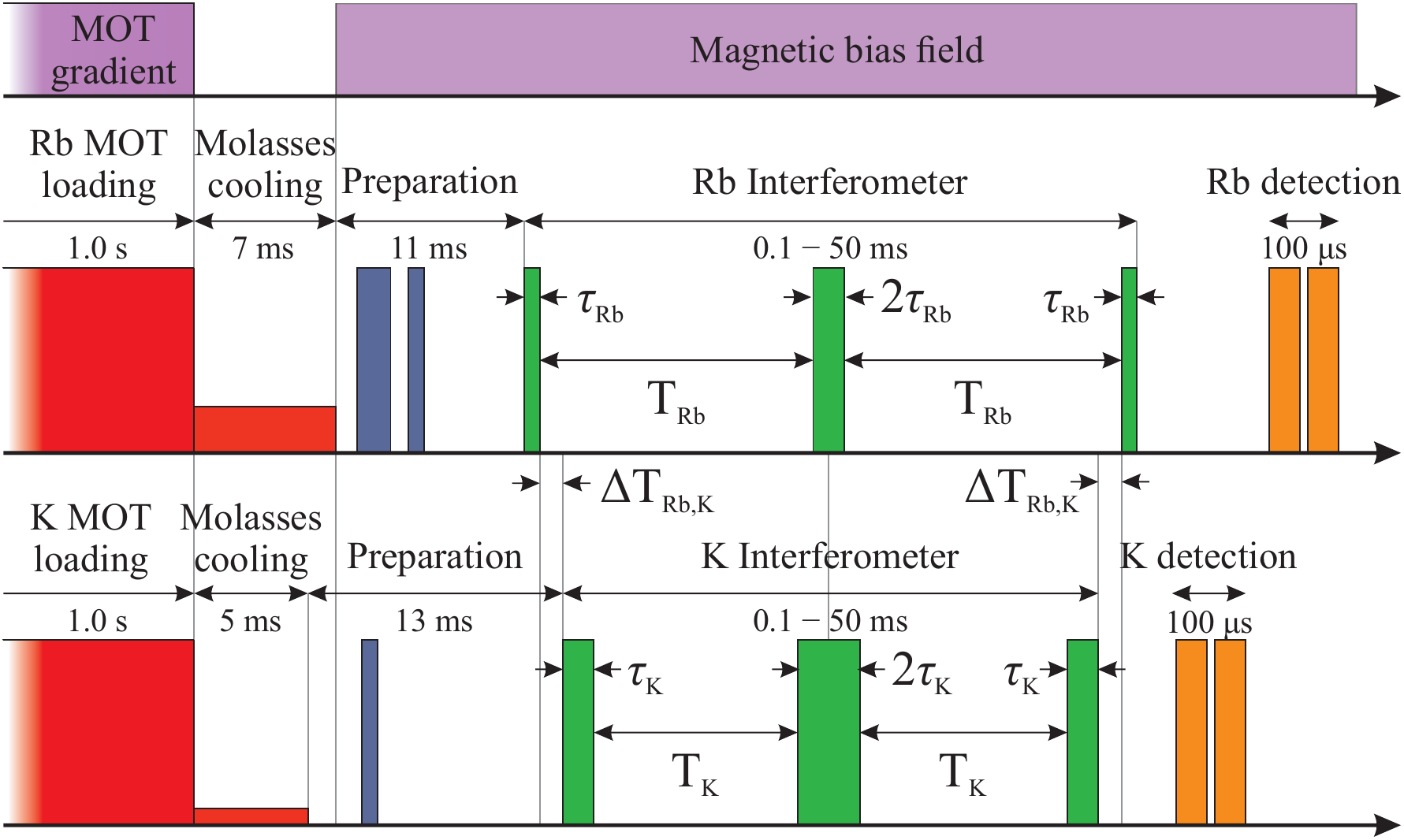}
  \caption{Symmetric timing sequence for the K-Rb interferometer. The first Raman pulse of the $^{39}$K interferometer is delayed by $\tau_{\rm Rb} + \Delta T_{\rm Rb, K}$ relative to that of $^{87}$Rb such that the central $\pi$-pulse of both interferometers occurs at the same time. The preparation stages (shown in blue) include both the internal state preparation pulses (microwave + push beam for $^{87}$Rb, depump for $^{39}$K), and the external state preparation via time-of-flight. Raman pulses are shown in green, and detection pulses in orange.}
  \label{fig:KRbInterferometerTiming}
\end{figure}
%--------------------------------------------

A typical experimental sequence for the K-Rb interferometer is shown in \Fig \ref{fig:KRbInterferometerTiming} and is carried out as follows. The MOT beams load approximately $2 \times 10^8$ ($7 \times 10^7$) atoms in 0.5 s, which is followed by a 7 ms (5 ms) molasses cooling stage for the $^{87}$Rb ($^{39}$K) sample. In addition to cooling, the rubidium molasses stage also pumps the atoms into the $\ket{F = 2}$ ground state. This is followed by a microwave $\pi$-pulse that transfers atoms into $\ket{F = 1, m_F = 0}$, and the remaining atoms are removed with a push beam resonant with the $F = 2$ to $F' = 3$ transition. During the potassium molasses, the frequency and intensity of the cooling and repump beams are modified in a similar manner to \Refs \cite{Landini2011, Gokhroo2011}. At the end of the molasses, the atoms are in a superposition of both hyperfine ground states, which is a critical part of the cooling mechanism for potassium \cite{Landini2011}. We detune our 767 nm push beam to the red of the $F = 2$ to $F' = 3$ transition by $\sim 17$ MHz (2.9 $\Gamma$) to optically pump the atoms into the $F = 1$ level with a 3 $\mu$s pulse. Following this depumping stage, the $^{39}$K atoms are distributed roughly equally amongst the magnetic sub-levels of the lower hyperfine ground state. With this system, we achieve temperatures of $\sim 3$ $\mu$K for $^{87}$Rb and $\sim 20$ $\mu$K for $^{39}$K, as confirmed by both time-of-flight imaging and velocity-sensitive Raman spectroscopy. After preparing the internal atomic states, we typically wait $\sim 12$ ms for the atoms to fall such that the Doppler resonance of both sets of counter-propagating Raman beams becomes non-degenerate. Additionally, we apply an external magnetic bias field between $1 - 2$ Gauss to shift the $\ket{F = 1, m_F = \pm 1}$ states of potassium away from the central $m_F = 0$ state on which we perform interferometry. The frequency of the Raman beams for both species is detuned by $-1.2$ GHz ($-200~\Gamma$) relative to the $F = 2$ to $F' = 3$ transition. We then apply the interferometry pulses in a symmetric fashion, such that the central $\pi$-pulse for both interferometers occurs at the same time, as shown in \Fig \ref{fig:KRbInterferometerTiming}. The delay between the $\pi/2$ pulses for either atom, $\Delta T_{\rm Rb, K}$, can be adjusted within the interrogation time of the rubidium interferometer, $T_{\rm Rb}$, in order to study correlations and effects related to the scale factor ratio, $\kappa$. Finally, we measure the atomic state populations for each atom via fluorescence detection on an avalanche photodiode (50 MHz bandwidth) within 100 $\mu$s of one another.

%================================================================================================
\section{Experimental results}
\label{sec:Results}

We now describe some experimental results obtained from the K-Rb interferometer. All of the data presented in this work were recorded in a laboratory environment, with the interferometer beams aligned along the vertical direction, and with no anti-vibration platform. To compensate for the Doppler shift due to gravity, the frequency difference between Raman beams for interferometers $j = 1 \equiv \rm{K}$ and $j = 2 \equiv \rm{Rb}$ is chirped at a rate of $\alpha_j \simeq k_j^{\rm eff} g$ to account for the gravity-induced Doppler shift of the falling atoms. This modifies the total phase shift of the interferometers from \Eqs \eqref{yj(a)}, $\Phi_j = S_j a + \phi_j$, to the following
\be
  \label{Phij}
  \Phi_j = S_j(a - \alpha_j/k_j^{\rm eff}) + \phi_j \simeq (k_j^{\rm eff} g - \alpha_j) T_j^2.
\ee
The last expression represents the case when both interferometers experience the same acceleration, $a_j = a = g$, and the scale factors can be approximated as $S_j \simeq k_j^{\rm eff} T_j^2$. Determining the location of the central fringe, for which $\alpha_j = k_j^{\rm eff} g$ is fixed for all $T_j$, yields a measurement of $g$. Using the fact that the sensitivity of the interferometer scales as $T_j^2$, absolute measurements of $g$ have been demonstrated at accuracies of a few $10^{-9}$ \cite{Peters2001, Farah2014, Schmidt2011, Hu2013}.

As discussed in the introduction, we are interested in measuring the differential acceleration $\Delta a$ between $^{39}$K and $^{87}$Rb. One way of achieving this is to measure the gravitationally-induced accelerations $g_{\rm K}$ and $g_{\rm Rb}$ from each interferometer independently by scanning the chirp rates, $\alpha_j$, in a low-noise environment. This is the approach recently employed for WEP tests with $^{39}$K and $^{87}$Rb by Schlippert \etal \cite{Schlippert2014}. However, at high levels of sensitivity (\ie large $T_j$), or in ``noisy'' environments, mirror vibrations can corrupt the fringes---making individual phase measurements more challenging. We now demonstrate the utility of the FRAC technique for measuring $g$ from a single interferometer under these conditions.

%--------------------------------------------
\begin{figure}[!tb]
  \centering
  \includegraphics[width=0.60\textwidth]{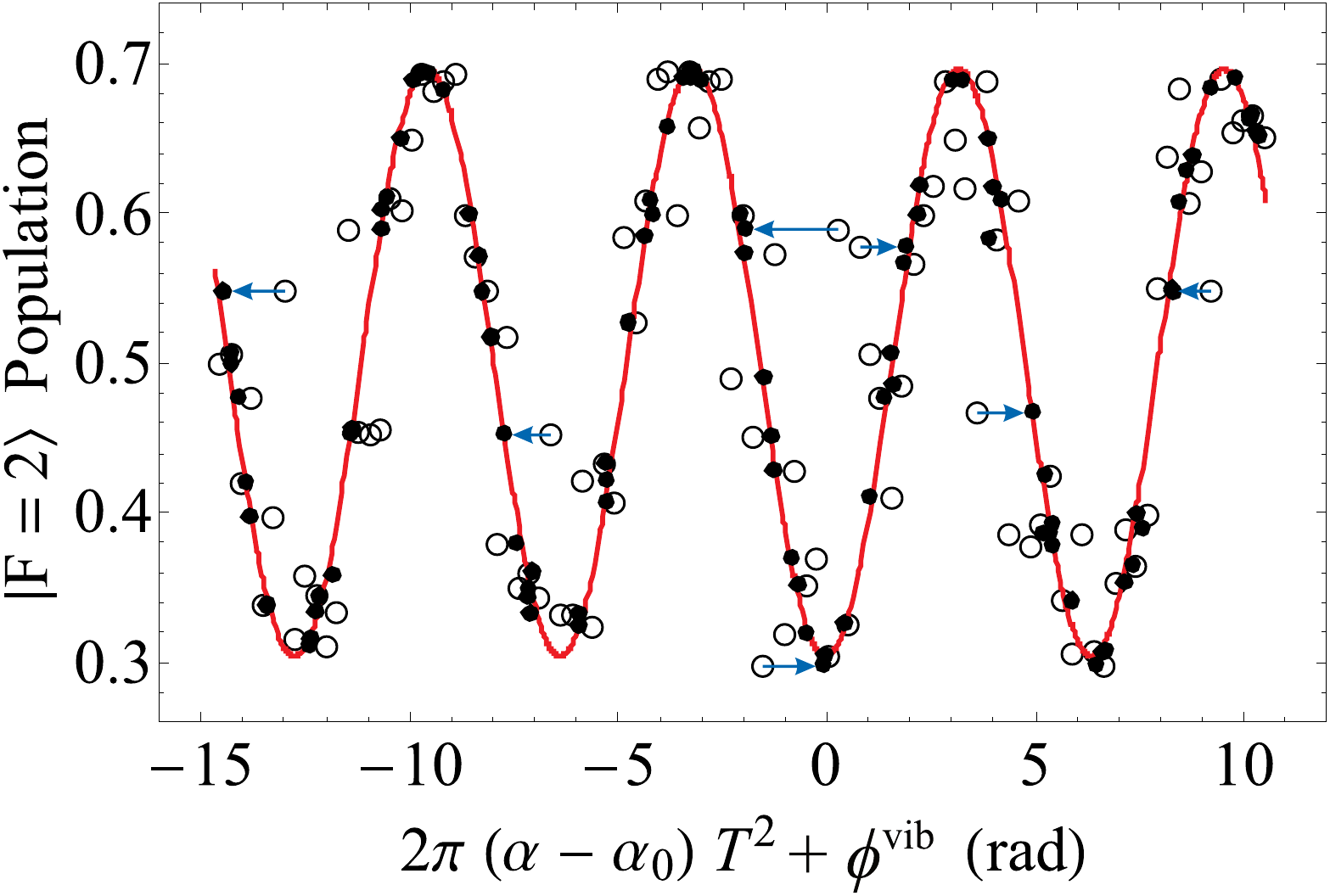}
  \caption{Interferometer fringes from the $^{87}$Rb interferometer operating at $T = 25$ ms at a contrast of $\sim 40\%$ without vibration isolation. The interferometer phase is scanned by both uncontrolled mirror motion, and by varying the chirp rate $\alpha$ between Raman beams about a central value of $\alpha_0 = 25.1355$ MHz/s. The open circles indicate raw measurements of the normalized atomic population in the $\ket{F = 2}$ state, while the closed circles are the same measurements after applying the FRAC phase correction, $\phi^{\rm vib}$, to each point. A few of these corrections are shown as blue arrows. The solid curve is a least-squares fit to the corrected data, resulting in a signal-to-noise ratio of $\sim 30$ and a relative statistical uncertainty of $10^{-7}$ in the determination of $g_{\rm Rb}$---corresponding to almost an order of magnitude improvement compared to the raw data.}
  \label{fig:RbFringeReconstruction}
\end{figure}
%--------------------------------------------

There are typically two approaches in which the FRAC method can be applied to restore the interference fringes of a single interferometer. The first approach is to let the interferometer phase be ``scanned'' randomly by vibrations while the laser-induced phase is held fixed. The reconstructed fringes in this case are purely a function of $\phi_j^{\rm vib}$, as shown in \Fig \ref{fig:AccelerometerCorrelation}. This mode of operation can be used to precisely calibrate the mechanical accelerometer by rescaling the voltage-to-acceleration sensitivity factor of the device such that the fringe period is $2\pi$\footnote{One advantage of performing this procedure is that the device can be precisely calibrated for the vibration spectrum on site. Depending on the bandwidth and spectral response of the device, the sensitivity can vary significantly with the vibration spectrum.}. The second approach is to scan the interferometer phase in a controlled manner, for example by varying the phase difference between Raman lasers, and to correct each phase using $\phi^{\rm vib}$ obtained during the same measurement interval. This procedure is illustrated in \Fig \ref{fig:RbFringeReconstruction}, where the fringes of a $T = 25$ ms $^{87}$Rb interferometer are shown before and after applying the FRAC correction. Here, the interferometer is operated without any vibration isolation in the presence of a root-mean-squared (rms) DC vibration noise of $a^{\rm vib}_{\rm rms} \simeq 6 \times 10^{-5}$ m/s$^2$ (integrated over the frequency response of the interferometer)---corresponding to an rms phase noise of $\phi^{\rm vib}_{\rm rms} = k_{\rm Rb}^{\rm eff} a^{\rm vib}_{\rm rms} T_{\rm Rb}^2 \simeq 0.6$ rad. Acceleration measurements were performed with a force-balance three-axis accelerometer (Nanometrics Titan, DC to 430 Hz bandwidth, 5 V/$g$ sensitivity). By applying the FRAC correction to these data, we improve the signal-to-noise ratio (SNR) and hence the uncertainty in the central fringe measurement by almost an order of magnitude. We estimate an individual phase correction uncertainty of $\delta \phi^{\rm vib} = 1/\mbox{SNR} \simeq 33$ mrad based on the improved SNR of $\sim 30$. With this method, we emphasize that the interferometer sensitivity is directly linked to the intrinsic noise of the accelerometer + signal acquisition system, and the quality of the coupling between the device and the Raman mirror. Therefore, modest improvements to any of these system components can result in a dramatic increase in the fringe SNR.

%------------------------------------------------------------------------------------------------
\subsection{K-Rb Interferometer Correlation}
\label{sec:KRbCorrelation}

Typically, when mirror motion is the dominant source of phase noise it is advantageous to use differential atom interferometry techniques to measure $\Delta a$ through the differential phase $\phi_d$. This requires a high level of correlation between interferometers in order to reject the common-mode phase noise. We now compare three methods of extracting $\phi_d$ from experimental data recorded in an environment with high vibrational noise, as in the case of onboard applications \cite{Geiger2011phd,Geiger2011}. These studies are also applicable to future high-sensitivity differential interferometers operated in low-noise environments \cite{Dickerson2013, Sugarbaker2013, Hartwig2015}.

%--------------------------------------------
\begin{figure*}[!tb]
  \centering
  \includegraphics[width=0.96\textwidth]{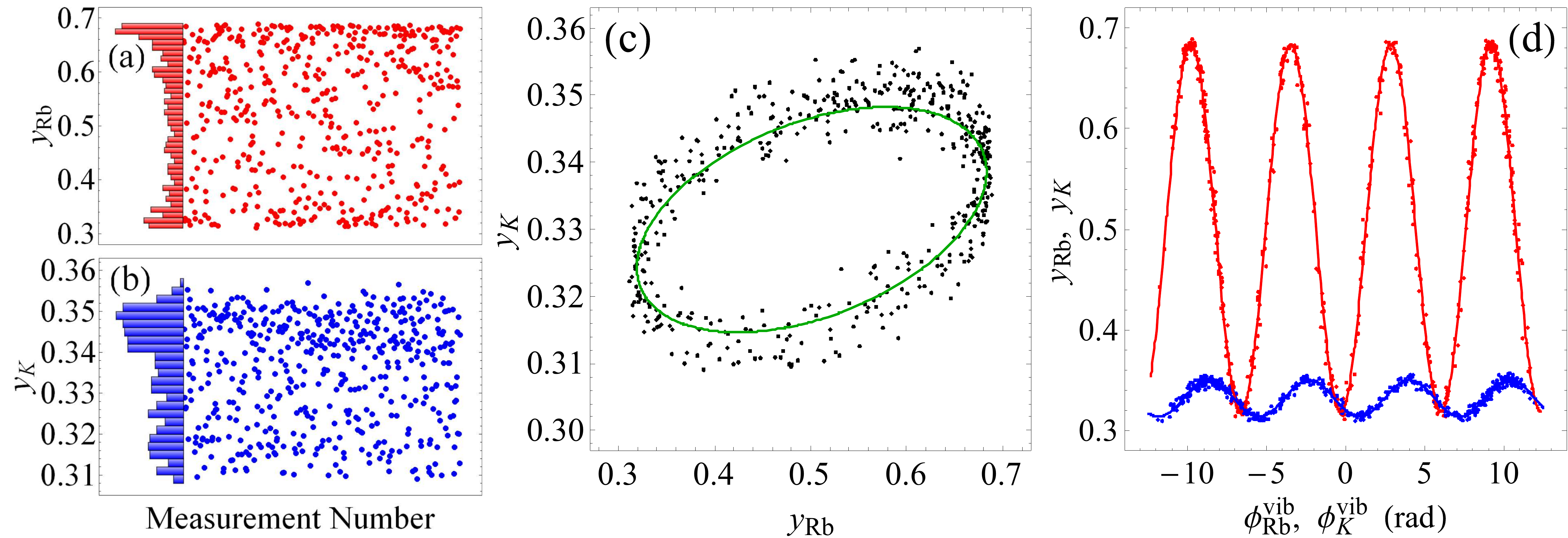}
  \caption{Measurements of normalized $\ket{F = 2}$ state populations from simultaneous K-Rb interferometers operating at $T \simeq 3$ ms. Fringes are scanned randomly by applying vibrational noise to the retro-reflection mirror. Graphs (a) and (b) show a time series of 500 measurements from the $^{87}$Rb and $^{39}$K interferometers, respectively, along with histograms of the populations. (c) Atomic populations from (a) and (b) plotted parametrically---indicating strong correlation between the two species. The solid green line is an ellipse fit to the data using the FGEF method, which yields a differential phase $\phi_d^{\rm ellipse} = 1.13(2)$ rad. A separate estimate from a Bayesian analysis gives $\phi_d^{\rm Bayes} = 1.18(2)$ rad. (d) Interferometer fringes reconstructed from measurements of mirror motion using the FRAC method. The red and blue curves correspond to least-squares fits to Rb and K data, respectively. The differential phase estimated from the fits is $\phi_d^{\rm FRAC} = 1.17(1)$ rad. Other interferometer parameters: pulse separations: $T_{\rm Rb} = 3.018$ ms, $T_{\rm K} = 3$ ms; $\pi/2$-pulse durations: $\tau_{\rm Rb} = 4$ $\mu$s, $\tau_{\rm K} = 6$ $\mu$s; delay between interferometers: $\Delta T_{\rm K, Rb} = 10$ $\mu$s; one-photon Raman detunings: $\Delta_{\rm Rb} = \Delta_{\rm K} \simeq -1.2$ GHz.}
  \label{fig:KRbFringes}
\end{figure*}
%--------------------------------------------

Figure \ref{fig:KRbFringes} shows data produced by quasi-simultaneous K-Rb interferometers at a total interrogation time of $2T = 6$ ms. Here, we held the chirp rate fixed at $\alpha_j \simeq k_j^{\rm eff} g$ for each species, and we applied strong vibrations to the system ($a^{\rm vib}_{\rm rms} \simeq 0.05$ m/s$^2$) such that the random vibration-induced phase $\phi_j^{\rm vib}$ spanned multiple fringes ($\phi^{\rm vib}_{\rm rms} \simeq 7.3$ rad). The vibrations were applied by mounting a heavy industrial fan on top of the support structure surrounding the vacuum system and running it during the experiment. Figure \ref{fig:KRbFringes}(a) shows a histogram of $^{87}$Rb $\ket{F = 2}$ population measurements, $y_{\rm Rb}$, which clearly indicates the characteristic bimodal probability distribution of a sinusoid. These distributions can be used to estimate the contrast, offset and SNR of the interferometer fringes as described in \Ref \cite{Geiger2011}. We note that the bimodal distribution is less pronounced for $^{39}$K in \Fig \ref{fig:KRbFringes}(b) owing to a smaller fringe contrast, and thus a lower SNR, compared to $^{87}$Rb. Despite this fact, the two sensors exhibit strong correlations, as confirmed by the ellipse in \Fig \ref{fig:KRbFringes}(c).

For these experimental parameters the scale factor ratio is $\kappa = S_{\rm K}/S_{\rm Rb} = 1.008$, and the Lissajous curve formed by parametrically plotting the atomic state populations, $y_{\rm Rb}$ and $y_{\rm K}$, is indistinguishable from an ellipse at the present level of offset noise. We measure a differential phase of $\phi_d^{\rm ellipse} = 1.13(2)$ rad from a least-squares fit to an ellipse using the FGEF method \cite{Szpak2012}. We also estimate $\phi_d^{\rm Bayes} = 1.18(2)$ rad using the Bayesian analysis described in \Sec \ref{sec:Bayesian} and \ref{app:Bayesian}. Here, it is worth mentioning that this non-zero differential phase does not originate from a WEP violation, but from systematic phase shifts in the experiment---primarily due to the quadratic Zeeman effect from an external magnetic bias field ($\sim 1$ G) that is used to sufficiently split the ground state magnetic sub-levels in $^{39}$K.

Figure \ref{fig:KRbFringes}(d) shows the output of each interferometer as a function of the vibration-induced phase, $\phi_j^{\rm vib}$. Here, the single-sensor fringes were reconstructed using the FRAC method using mirror vibration measurements from a broadband micro-electro-mechanical accelerometer (Colibrys SF3600, DC to 1 kHz bandwidth, 1.2 V/$g$ sensitivity). From these data the differential phase shift between interferometers is clearly visible. Sinusoidal least-squares fits to each fringe yield $\phi_d^{\rm FRAC} = \phi_{\rm K} - \kappa \phi_{\rm Rb} = 1.17(1)$ rad. Here, the statistical uncertainty $\delta \phi_d^{\rm FRAC}$ was computed from the quadrature sum of each interferometer phase error. The value of $\phi_d$ estimated from the Bayesian analysis and the FRAC method are in good agreement. On the other hand, the differential phase from the ellipse fit is underestimated by $\sim 40$ mrad, \ie $2\sigma$ below $\phi_d^{\rm Bayes}$ and $\phi_d^{\rm FRAC}$. We attribute this discrepancy to the inherent bias of ellipse-fitting techniques (see \ref{app:Ellipse}), which increases with the level of offset noise or differential phase noise in either interferometer.

We emphasize that a crucial input parameter for the Bayesian analysis is the common phase range. We use the accelerometer data to estimate this range once the experiment is complete: $\phi_c \in [\min(\phi^{\rm vib}_{\rm Rb}), \max(\phi^{\rm vib}_{\rm Rb})]$. However, if an accelerometer is not available, it is also possible to estimate this range using the raw data from a single interferometer. For example, one can reduce the interrogation time until the sensitivity to vibrations reaches a point where interference fringes are clearly visible. By measuring the rms scatter of the phase about a reference sinusoid, one can estimate the level of vibration noise via the relation $a^{\rm vib}_{\rm rms} = \phi^{\rm vib}_{\rm rms}/S$. Once $a^{\rm vib}_{\rm rms}$ is known, this relation can be inverted to determine the range of phase scanned by the same level of vibrations at larger sensitivities/interrogation times.

The data shown in \Fig \ref{fig:KRbFringes}(d) also indicate that the combined differential-atomic-sensor + mechanical-accelerometer system is capable of efficiently rejecting common vibrational noise. We estimate a rejection factor of $\gamma = k^{\rm eff} a^{\rm vib}_{\rm rms} T^2 / \delta \phi_d^{\rm FRAC} \simeq 730$ for these data.

%--------------------------------------------
\begin{figure}[!tb]
  \centering
  \includegraphics[width=0.90\textwidth]{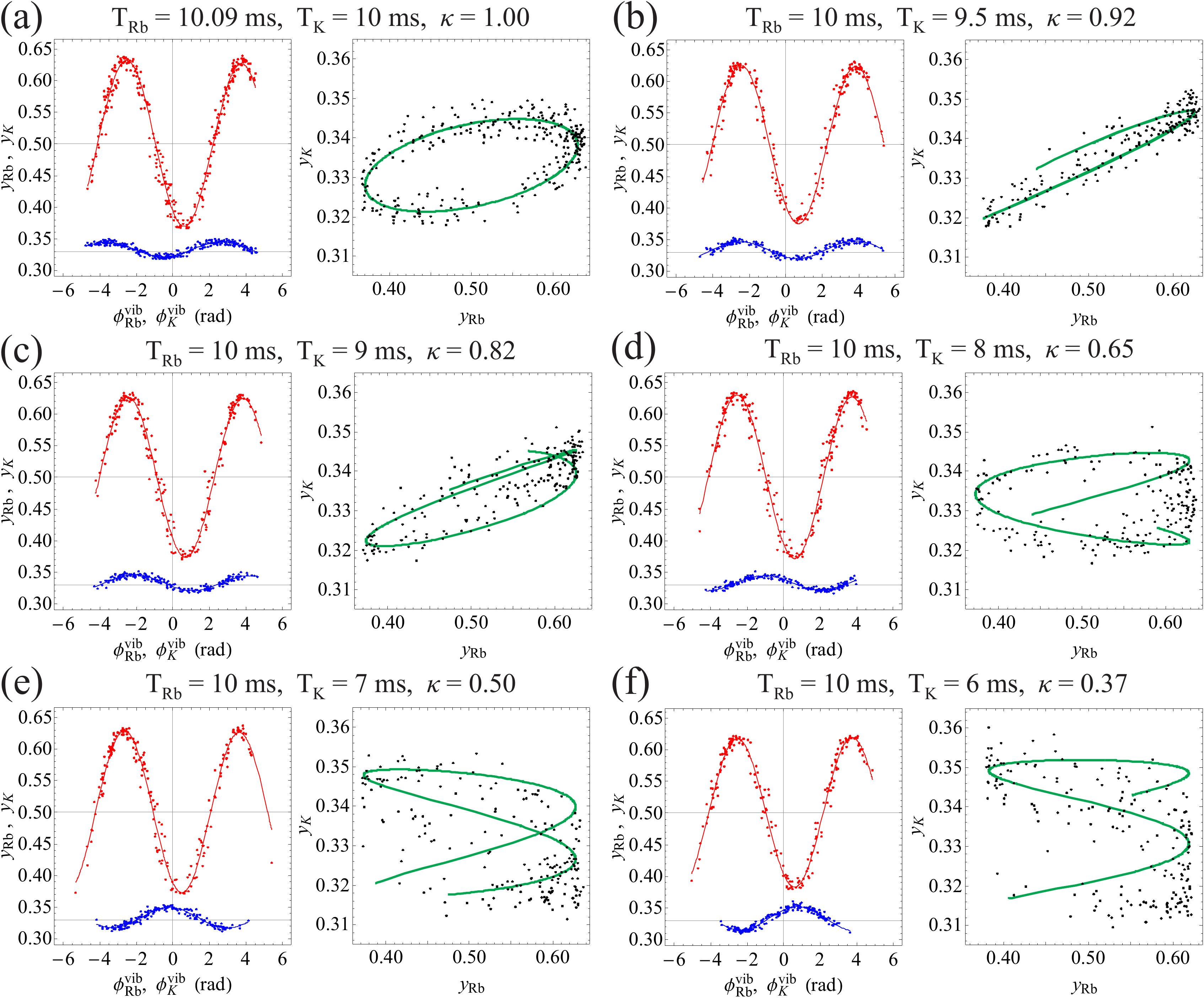}
  \caption{Correlation studies for different levels of temporal overlap. The symmetric, quasi-simultaneous K-Rb interferometer was operated with $T_{\rm Rb} = 10$ ms and the interrogation time for potassium was varied between $T_{\rm K} = 6 - 10$ ms. The interferometer phase was scanned by externally applied vibrations and individual fringes were restored using the FRAC method. Parametric plots of the atomic populations are shown to the right, along with the expected Lissajous curve (solid green line). These curves result from plotting the fit functions to each reconstructed fringe parametrically. There is a clear disagreement between the predicted Lissajous curves and the data for $T_{\rm Rb} - T_{\rm K} \gtrsim 2$ ms. Other interferometer parameters: $\tau_{\rm Rb} = \tau_{\rm K} = 3$ $\mu$s; $\Delta_{\rm Rb} = \Delta_{\rm K} \simeq -1.2$ GHz.}
  \label{fig:LissajousFigures}
\end{figure}
%--------------------------------------------

Figure \ref{fig:LissajousFigures} displays the results of a correlation study between rubidium and potassium interferometers operating at a total interrogation time of $2T = 20$ ms. Similar to \Fig \ref{fig:KRbFringes}, the interferometer phases are scanned by externally applied vibrations ($a_{\rm rms}^{\rm vib} \simeq 1.7 \times 10^{-3}$ m/s$^2$, $\phi_{\rm rms}^{\rm vib} \simeq 2.7$ rad at $T_{\rm Rb} = T_{\rm K} = 10$ ms). The vibrations were applied using the same method as previously mentioned, but with the fan set on a slower rotation setting. Here, we vary the interrogation time of potassium, $T_{\rm K}$, in a symmetric way with respect to rubidium such that the centers of the $\pi$-pulses coincide. This optimizes the degree to which the vibration-induced phase noise remains common-mode, while modifying the degree of temporal overlap between interferometers. It also allows us to control the scale factor ratio since $\kappa$ scales at $(T_{\rm K}/T_{\rm Rb})^2$.

From \Fig \ref{fig:LissajousFigures}, three features are clearly visible as $T_{\rm K}$ is decreased. First, the potassium fringes undergo a phase shift that modifies the differential phase relative to the rubidium fringes. This feature, along with the fact that the scale factor ratio is varied, causes the shape of the Lissajous figures to change, as shown by the solid green curves. Second, the phase range scanned by the potassium interferometer reduces, since it scales as $T_{\rm K}^2$. Finally, the level of correlation between the interferometers degrades as the temporal overlap decreases. This is evident from the lack of agreement between the data and the predicted Lissajous curves, particularly for $T_{\rm K} \lesssim 8$ ms.

Regardless of this degradation of correlation and temporal overlap between interferometers, the differential FRAC method is able to restore the interference fringes with a good SNR ($\sim 30$ for $^{87}$Rb, $\sim 10$ for $^{39}$K, limited by uncorrelated offset noise). The vibration rejection factor for each of the data sets shown in \Fig \ref{fig:LissajousFigures} is approximately $\gamma \sim 100$. This permits unbiased estimates of $\phi_d$ with a statistical uncertainty at the level of $\delta \phi_d \sim 25$ mrad with 300 points---corresponding to WEP test with a statistical sensitivity of $\delta \eta \simeq \delta \phi_d/k^{\rm eff}_{\rm Rb} g T_{\rm Rb}^2 = 1.6 \times 10^{-6}$ per data set. The robustness of the differential FRAC technique under ``noisy'' conditions makes it an ideal candidate for future WEP tests \cite{Altschul2015, Schubert2013, Aguilera2014}, or other differential atom interferometry applications \cite{Rosi2015, Canuel2014}.

In contrast, for Bayesian estimation, an increase in uncorrelated phase noise is problematic. When the ``common phase'' becomes largely uncorrelated, the Bayesian method can converge on multiple possible $\phi_d$, or may not converge at all. For these data, we find that by $T_{\rm K} = 9$ ms the Bayesian estimate of $\phi_d$ is not consistent with the FRAC estimate, and for $T_{\rm K} \lesssim 8$ ms the analysis is not able to converge on a unique value. We note that these particular results are strongly dependent on the level of phase noise, the degree of temporal overlap, the value of $\phi_d$ and the scale factor of each interferometer. In the following section, we study some of these dependencies more quantitatively.

%------------------------------------------------------------------------------------------------
\subsection{Comparison of Bayesian and FRAC methods as a function of $\kappa$ and $\phi_d$}
\label{sec:Comparison}

We have tested the functionality and accuracy of both the Bayesian and FRAC methods for extracting $\phi_d$ from experimental data acquired under various conditions. Specifically, we are interested in the accuracy of these techniques over (i) the full range of differential phase $\phi_d \in [0, \pi]$, and (ii) a broad range of interferometer scale factor ratios $\kappa = S_{\rm K}/S_{\rm Rb}$. To investigate these two aspects, we recorded data using the symmetric interferometer configuration shown in \Fig \ref{fig:KRbInterferometerTiming} with different interrogation times, $T_{\rm Rb}$ and $T_{\rm K}$. Since $\kappa$ is proportional to $(T_{\rm K}/T_{\rm Rb})^2$, each configuration of $T_j$ corresponds to a different scale factor ratio. Additionally, the differential phase is modified with each $T_{\rm K}$ due to a systematic phase shift of the potassium interferometer from an external magnetic field. Therefore, we are able to study both effects with a single data set.

%--------------------------------------------
\begin{figure}[!tb]
  \centering
  \subfigure{\includegraphics[width=0.40\textwidth]{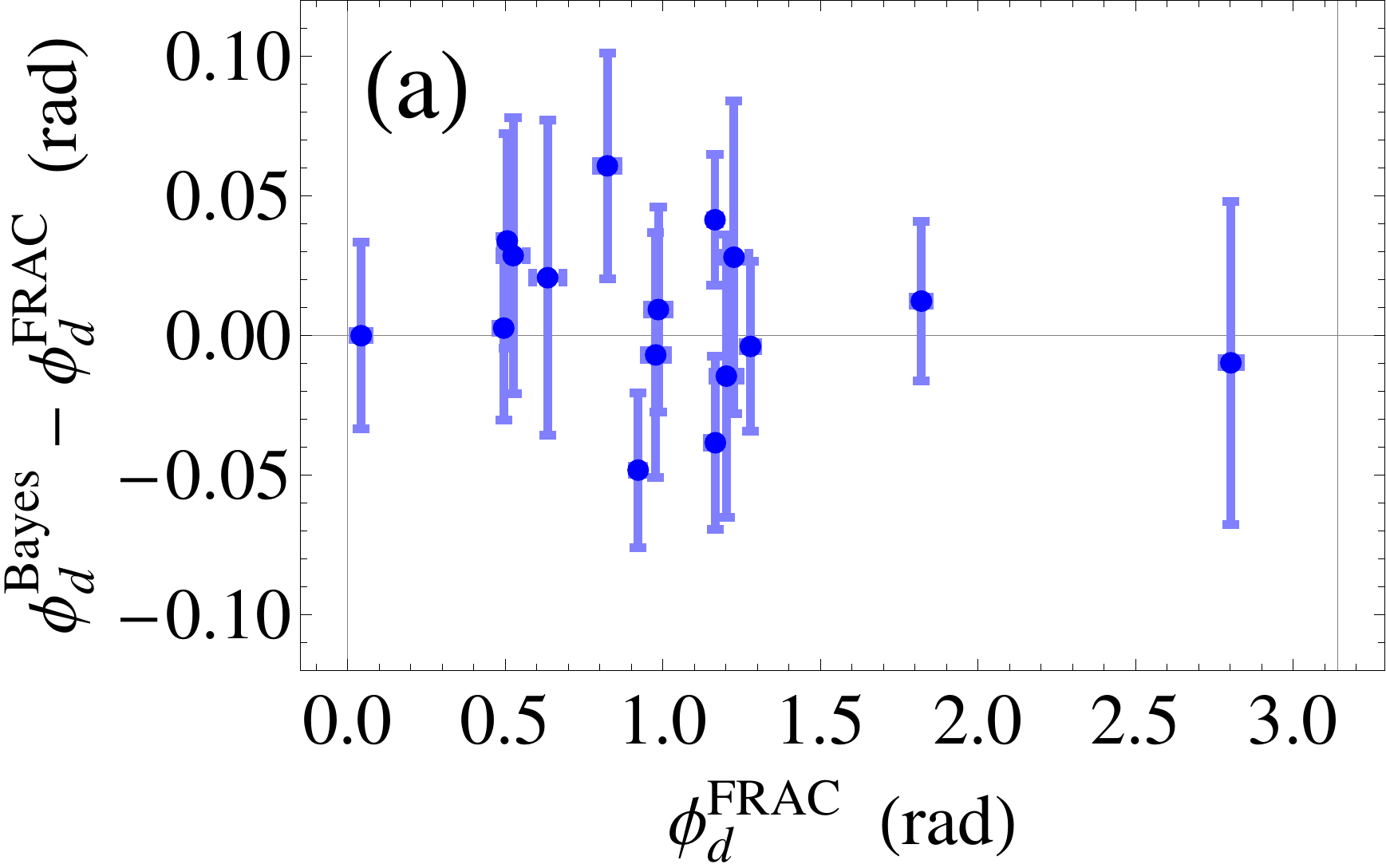}}
  \hspace{0.2cm}
  \subfigure{\includegraphics[width=0.40\textwidth]{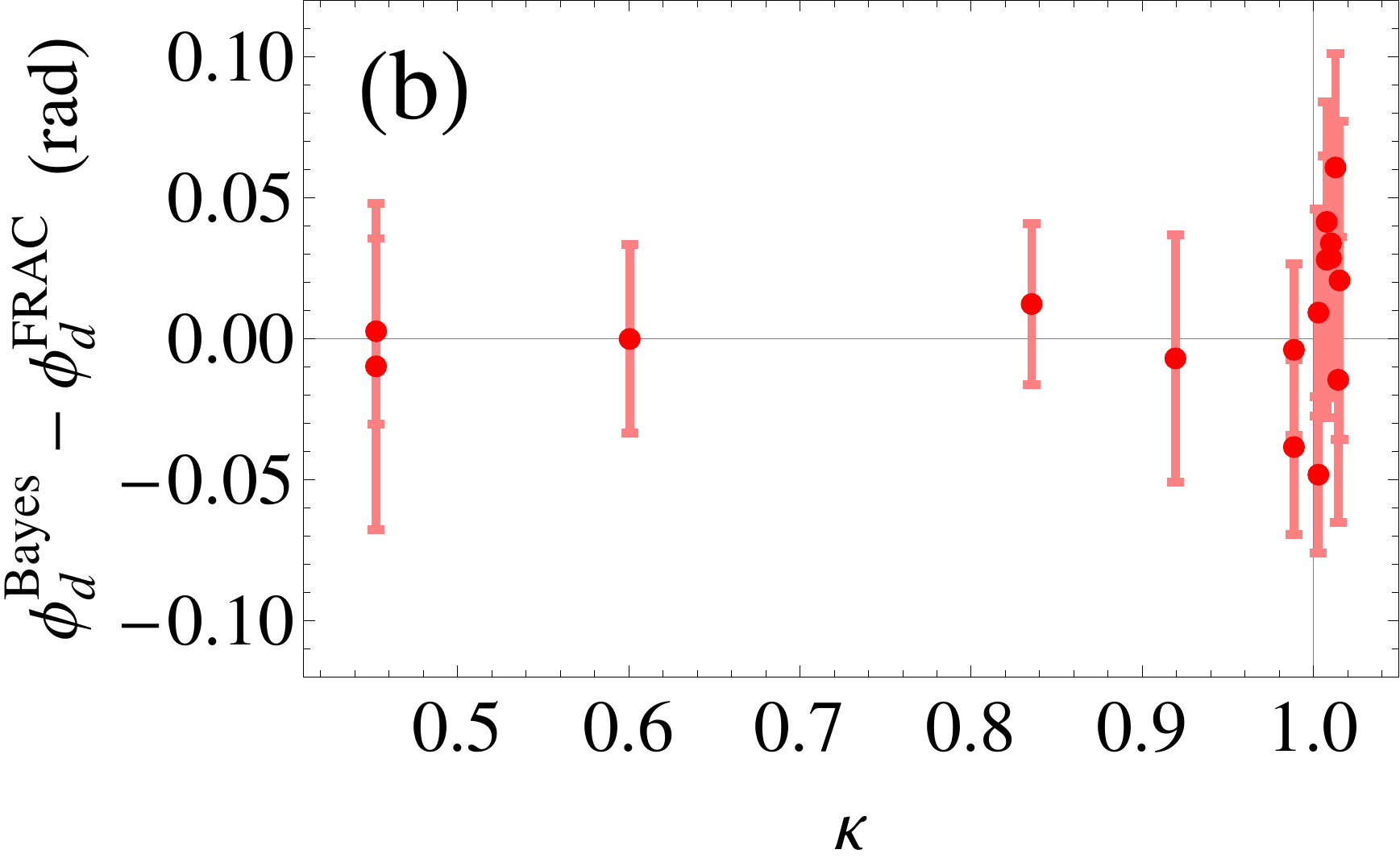}}
  \caption{Comparison of differential phase estimates from the Bayesian and FRAC methods for various $\phi_d$ and $\kappa$. (a) Difference between Bayesian and FRAC estimates $\phi_d^{\rm Bayes} - \phi_d^{\rm FRAC}$ as a function $\phi_d^{\rm FRAC} = \phi_{\rm K} - \kappa \phi_{\rm Rb}$. The vertical error bars represent the combined statistical uncertainty of both estimates. (b) Difference between Bayesian and FRAC estimates as a function of $\kappa = S_{\rm K}/S_{\rm Rb}$ corresponding to each point in (a). The majority of the data points shown in both figures are consistent with zero to within $1\sigma$ of uncertainty, indicating agreement between the two techniques at the level of $\sim 40$ mrad. The standard deviation of offset noise for each data set was typically $\sigma_{B_{\rm K}} \simeq 0.20$ and $\sigma_{B_{\rm Rb}} \simeq 0.05$ in the normalized space ($n_1$ and $n_2$). A value of $\sigma_{\phi_d} = 0.05$ rad was used for the differential phase noise of all data sets. The range of common phase noise was estimated from accelerometer measurements.}
  \label{fig:DiffPhaseEstimation}
\end{figure}
%--------------------------------------------

Figure \ref{fig:DiffPhaseEstimation} shows a comparison between Bayesian and FRAC estimates of $\phi_d$, using the FRAC estimate as a reference. We varied $T_{\rm Rb}$ from 1 to 5 ms, and $T_{\rm K}$ independently in the vicinity of $T_{\rm Rb}$ such that the scale factor ratio was modified over a relatively broad range ($\kappa \simeq 0.45$ to 1.01). The phase noise due to the externally applied vibrations was kept quasi-common-mode between sensors by ensuring that $T_{\rm K}$ was within a few 100 $\mu$s of $T_{\rm Rb}$. Over this range of $T_{\rm Rb}$ and $T_{\rm K}$, we found that the differential phase ranged from roughly $\phi_d = 0$ to 2.8 rad as a result of a systematic shift of the potassium interferometer. It is clear from \Fig \ref{fig:DiffPhaseEstimation} that there is a high degree of correlation between the Bayesian and FRAC estimates, which is consistent with our expectations based on the simulations discussed in \Sec \ref{sec:Bayesian}. The error bars in this figure were computed from the combined statistical uncertainties of both methods, which both typically yield $\delta \phi_d \sim 30$ mrad at the present level of noise.

To summarize, we find that the difference between the two estimates is consistent with zero within a typical total uncertainty of $\sim 40$ mrad. These data confirm that the two analysis techniques produce unbiased estimates of $\phi_d$ for dual-species interferometers with vastly different scale factors. We discuss further the advantages and limitations of these two techniques in the following section.

%================================================================================================
\section{Advantages and limitations of the methods}
\label{sec:Limitations}

As discussed in \Sec \ref{sec:Bayesian}, Bayes' method is optimally efficient and yields a statistical error that scales as $1/\sqrt{N}$, compared to more heuristic fitting techniques which converge more slowly. This improved efficiency is a clear advantage of the Bayesian estimator compared to the FRAC analysis. However, the disadvantage is that it requires a priori information about the system, such as noise levels and interferometer contrasts, and it requires significant computational resources to evaluate. Furthermore, it is only a viable solution for simultaneous interferometer configurations that exhibit a high-degree of phase correlation.

In contrast to the Bayesian estimator, the FRAC method requires only the interferometer timing parameters and a sensitive accelerometer that is well-coupled to the reference mirror in order to function accurately. It does not assume any particular interferometer configuration or require any additional system information. The FRAC method also has applications in absolute interferometry, as has been previously demonstrated in \Refs \cite{Gillot2014, LeGouet2008, Merlet2009, Farah2014, Geiger2011}. Additionally, it is fast enough to be used for real-time feedback, which has been shown to improve single-sensor sensitivity \cite{Lautier2014}.

%--------------------------------------------------
\setlength{\tabcolsep}{4pt}
\begin{table}[!tb]
  \centering
  \small
  \begin{tabular}{|c|c|cccc|cccc|}
    \hline
    \multicolumn{2}{|c|}{}            & \multicolumn{4}{|c|}{Single Sensor}          & \multicolumn{4}{|c|}{Differential Sensor} \\
    \hline
                     & Noise          & \multicolumn{4}{|c|}{Frequency Range (Hz)}   & \multicolumn{4}{|c|}{Frequency Range (Hz)} \\
    $T_{\rm Rb}$ (s) & Source         & DC~$-~1$ & $1-10$ & $10-100$ & DC~$-~\infty$ & DC~$-~1$ & $1-10$ & $10-100$ & DC~$-~\infty$ \\
    \hline
    \hline
    \multirow{2}{*}{0.01}
    & $(\phi_{\rm rms}^{\rm vib})$    & 0.023 & 0.779 & 1.071 & 1.324 & 0.000 & 0.000 & 0.004 & 0.004 \\
    & $(\phi_{\rm rms}^{\rm self})$   & 0.000 & 0.002 & 0.002 & 0.003 & 0.000 & 0.002 & 0.002 & 0.003 \\
    \hline
    \multirow{2}{*}{0.1}
    & $(\phi_{\rm rms}^{\rm vib})$    & 2.240 & 19.62 & 2.962 & 19.97 & 0.001 & 0.380 & 0.140 & 0.405 \\
    & $(\phi_{\rm rms}^{\rm self})$   & 0.050 & 0.078 & 0.005 & 0.093 & 0.050 & 0.078 & 0.005 & 0.093 \\
    \hline
    \multirow{4}{*}{1}
    & $(\phi_{\rm rms}^{\rm vib})$    & 58.13 & 23.42 & 3.779 & 62.79 & 1.067 & 2.556 & 1.511 & 3.155 \\
    & $(\phi_{\rm rms}^{\rm self})$   & 2.936 & 0.160 & 0.006 & 2.940 & 2.936 & 0.160 & 0.006 & 2.940 \\
    & $(\phi_{\rm rms}^{\rm vib})^*$  & 28.06 & 1.868 & 0.021 & 28.12 & 0.303 & 0.098 & 0.009 & 0.319 \\
    & $(\phi_{\rm rms}^{\rm self})^*$ & 0.294 & 0.016 & 0.001 & 0.294 & 0.294 & 0.016 & 0.001 & 0.294 \\
    \hline
  \end{tabular}
  \caption{Comparison between phase noise for a single interferometer, and two coupled interferometers with effective wave vectors $k^{\rm eff}_{\rm Rb}$ and $k^{\rm eff}_{\rm K}$. The rms phase noise (in radians) due to vibrations ($\phi^{\rm vib}_{\rm rms}$) and the self-noise of the mechanical accelerometer ($\phi^{\rm self}_{\rm rms}$) are shown for different frequency bands and interrogation times, $T_{\rm Rb}$. The noise from each band is summed in quadrature to obtain the total noise. Contributions less than 1 mrad are not shown. For the simultaneous differential sensor, it is assumed that $k^{\rm eff}_{\rm Rb} T_{\rm Rb}^2 = k^{\rm eff}_{\rm K} T_{\rm K}^2$. The rms phase noise $\phi_{\rm rms}^{\rm vib}$ was computed from \Eq \eqref{(phi_rms^vib)^2} using model \eqref{Sa(omega)} for the power spectral density $S_a(\omega)$ of ground accelerations in a ``quiet'' location \cite{LeGouet2008, Merlet2009} with integrated rms noise $1.4 \times 10^{-4}~g$ [see \Fig \ref{fig:Hj(omega)-Sa(omega)}(b)]. The quantity $\phi_{\rm rms}^{\rm self}$ was computed in a similar manner by replacing $S_a(\omega)$ with the self-noise spectrum of the accelerometer---here assumed to be white noise with $|\mathcal{S}_a|^{1/2} \simeq 3.2 \times 10^{-8}$ $g/\sqrt{\rm Hz}$. Quantities in the last row indicated by ``*'' correspond to low-noise conditions that can be achieved with passive vibration isolation (integrated rms noise $1.4 \times 10^{-6}~g$), and an accelerometer with 10 times smaller self-noise of $|\mathcal{S}_a|^{1/2} \simeq 3.2 \times 10^{-9}$ $g/\sqrt{\rm Hz}$.}
  \label{tab:SensitivityLimits}
\end{table}
%--------------------------------------------------

Table \ref{tab:SensitivityLimits} contains estimates of the phase noise for a single $^{87}$Rb interferometer, and two coupled interferometers of $^{87}$Rb and $^{39}$K. The rms phase spread due to vibration noise ($\phi^{\rm vib}_{\rm rms}$) and the self-noise of the accelerometer ($\phi^{\rm self}_{\rm rms}$) are shown for various frequency bands and interrogation times. For a single sensor analyzed with the FRAC method, the vibration-induced noise $\phi^{\rm vib}_{\rm rms}$ represents the spread of phase on the uncorrected fringes, while the quantity $\phi^{\rm self}_{\rm rms}$ indicates the residual phase noise present on the corrected fringes. Since this term is directly linked to the intrinsic noise of the mechanical accelerometer, it represents a fundamental limitation of the standard FRAC method. To give a quantitative example, based in the self-noise of the Titan accelerometer used in our experiments ($3.2 \times 10^{-8}$ $g/\sqrt{\rm Hz}$), the corresponding phase noise reaches $\sim 90$ mrad for an interrogation time of 100 ms, and $\sim 3$ rad by $T_{\rm Rb} = 1$ s. With this level of self-noise, fringes cannot be reconstructed accurately. However, for a state-of-the-art device with an order of magnitude smaller self-noise ($3.2 \times 10^{-9}$ $g/\sqrt{\rm Hz}$), the phase noise decreases by a further factor of 10---allowing a least-squares fit to accurately converge on the fringe phase. We also point out that the noise contributions from both vibrations and self-noise are smallest at high frequencies---a result of the natural low-pass filtering of atom interferometers. Thus, high-bandwidth accelerometers are generally not required to implement the FRAC method with a single sensor. For instance, for ground-based WEP test facilities targeting a few $10^{-15}$ \cite{Dimopoulos2007, Dickerson2013, Hartwig2015} using interrogation times of $T \sim 1$ s, we estimate that a mechanical accelerometer with a self-noise less than $10^{-11}$ $g/\sqrt{\rm Hz}$ in the DC to 1 Hz frequency band will yield a phase noise contribution below the projected shot-noise limit of $\sim 1$ mrad for each interferometer.

When employing the \emph{differential} FRAC method with two simultaneous interferometers, the self-noise of the accelerometer contributes to the phase of both sensors. Thus, in table \ref{tab:SensitivityLimits} (where $\kappa = 1$ and $T_{\rm K} = T_{\rm Rb} \sqrt{k_{\rm Rb}^{\rm eff}/k_{\rm K}^{\rm eff}}$), we have indicated the same values for $\phi_{\rm rms}^{\rm self}$ in the corresponding columns for both single and differential sensors. However, we emphasize that one can measure the differential phase significantly more accurately than the self-noise limit for a single sensor. This is because the noise introduced by the accelerometer is correlated between the two interferometers---reducing the uncertainty in the determination of $\phi_d$, as discussed in \Sec \ref{sec:FRAC}. A recent study \cite{PereiraDosSantos2015} has shown that uncertainties close to the quantum-projection-noise limit can be obtained with this method when the interferometers are in phase ($\phi_d = 0$) and the accelerometer exhibits a conservative level of self-noise ($\phi_{\rm rms}^{\rm self} \lesssim 0.3$ rad). In general, for the differential FRAC method to function well for all values of $\phi_d$ and $\kappa$, the self-noise of the accelerometer should correspond to less than $\pi/2$ in phase noise for each sensor---allowing individual fringes to be accurately fit. However, for the special case of $\phi_d = 0$ and $\kappa = 1$, the requirements on the accelerometer noise are much less stringent. For instance, \Ref \cite{PereiraDosSantos2015} indicates that reliable fits can be obtained with up to $\phi_{\rm rms}^{\rm self} \sim 20$ rad. For these reasons, we emphasize that state-of-the-art mechanical accelerometers are not required to make sensitive measurements of $\phi_d$ with long-baseline differential interferometers. We anticipate that competitive levels of accuracy can be achieved with compact devices that feature a moderate level of sensitivity.

For two coupled interferometers exhibiting different wave vectors, the vibration-induced phase noise is not identical and thus cannot be perfectly rejected at all frequencies. The values of $\phi^{\rm vib}_{\rm rms}$ listed in the last four columns of table \ref{tab:SensitivityLimits} contribute directly to $\phi_d$---representing the level of uncorrelated \emph{differential} phase noise in the system. We estimate that by $T_{\rm Rb} = 1$ s the differential phase noise reaches a level of $\sim 3$ rad. However, we note that the differential transfer function [\Eq \eqref{Hd(omega)}] rejects most efficiently at frequencies below $\sim 1/T$, and this estimate is directly linked to the vibration spectrum used. In a quieter environment, such as that achieved with a vibration isolation platform \cite{LeGouet2008, Merlet2009} or in a satellite \cite{Tino2013}, the phase noise can be reduced by an order of magnitude or more. At this point, the Bayesian method can be employed---which easily handles differential phase noise. Since the sensitivity scales as $\phi^{\rm vib}_{\rm rms}/\sqrt{N}$, the analysis simply requires more measurements for larger $\phi^{\rm vib}_{\rm rms}$ to reach a given level of precision.

%================================================================================================
\section{Conclusion}
\label{sec:Conclusion}

We have described and demonstrated experimentally two new analysis techniques for extracting the differential phase from coupled atom interferometers with different scale factors, $S_j$. A non-unity ratio $\kappa = S_1/S_2$ can result from using atoms with different $k_j^{\rm eff}$, or from interferometers with different interrogation times, $T_j$. We also carried out correlated phase measurements between simultaneous interferometers of two chemical elements exhibiting different scale factors, and we have demonstrated a vibration rejection factor as large as $\gamma \simeq 730$. This system was used to validate the Bayesian and FRAC analysis methods, as well as a new ellipse fitting procedure \cite{Szpak2012}, for extracting $\phi_d$. Furthermore, the FRAC method was used to demonstrate a statistical sensitivity for the E\"{o}tv\"{o}s parameter of $\delta \eta = 1.6 \times 10^{-6}$ per measurement with $T = 10$ ms interferometers operating in a ``noisy'' environment.

Presently, the contrast and SNR of our $^{39}$K interferometer fringes are limited by, respectively, the temperature of the atoms and technical noise present in the slave laser used for cooling, interferometry and detection. The latter issue results in uncorrelated offset noise, which reduces the correlation between interferometers. This, in turn, increases the error on $\phi_d$, which degrades the vibration rejection factor $\gamma$ and the sensitivity $\delta \eta$. We anticipate that a modest reduction of both the technical noise, and further cooling the $^{39}$K sample in a gray molasses \cite{Nath2013, Salomon2013}, will result in a substantial improvement in the correlation between rubidium and potassium. A precise determination of $\eta$ with our apparatus, including a complete evaluation of systematic effects, is beyond the scope of this work, but will be the subject of a future publication.

Both the generalized Bayesian and differential FRAC methods yield unbiased estimates of $\phi_d$ for any scale factor ratio, $\kappa$, and are robust against experimental parameters such as the common phase range scanned by the two interferometers, or the level of uncorrelated offset noise present in the system. These features make both methods ideal for applications of dual-species interferometry where, until now, the available analysis tools could accommodate only systems that exhibit either $\kappa = 1$ or low levels of common phase noise. These new methods are also appealing for gradiometer configurations using the same atoms \emph{and} the same $T_j$ \cite{PereiraDosSantos2015}, which have previously been utilized for precisely measuring the gravitational constant $G$ and gravity gradients \cite{Fixler2007, Rosi2014, Biedermann2015, McGuirk2002, Wu2009, Sorrentino2014, Lamporesi2008}.

The freedom to vary the scale factor, the interrogation time or phase of either interferometer independently can be advantageous for studying systematic effects, interactions between atomic species \cite{Kuhn2014}, or for shifting the differential phase toward a region of higher sensitivity. Examples of such regions include $\phi_d = \pi/2$ in the case of ellipse-fitting methods, and $\phi_d = 0$ or $\pi$ for the FRAC technique \cite{PereiraDosSantos2015}. Both the FRAC and Bayesian methods also eliminate the systematic shift introduced on the measurement of $\Delta a$ when using dual-species interferometers with $\kappa \not= 1$---making them well-suited for upcoming WEP tests on ground \cite{Dimopoulos2007, Dickerson2013, Sugarbaker2013, Hartwig2015}, in microgravity \cite{Barrett2014, Vogel2006, Varoquaux2009, VanZoest2010, Muntinga2013}, and in Space \cite{Altschul2015, Sorrentino2011, Tino2013, Schubert2013, Aguilera2014}.

%================================================================================================
%\acknowledgments
\section*{Acknowledgements}

This work is supported by the French national agencies CNES (Centre National d'Etudes Spatiales), l'Agence Nationale pour la Recherche, the D\'{e}l\'egation G\'{e}n\'{e}rale de l'Armement, the European Space Agency, IFRAF (Institut Francilien de Recherche sur les Atomes Froids), action sp\'{e}cifique GRAM (Gravitation, Relativit\'{e}, Astronomie et M\'{e}trologie) and RTRA ``Triangle de la Physique''. B. Barrett and L. Antoni-Micollier thank CNES and IOGS for financial support. P. Bouyer thanks Conseil R\'{e}gional d'Aquitaine for the Excellence Chair. Finally, the ICE team would like to thank the following people: F. Pereira Dos Santos of the laboratory SYRTE and V. M\'{e}noret of MuQuans for helpful discussions; D. Holleville, B. Venon, F. Cornu of SYRTE and J.-P. Aoustin of the laboratory GEPI for their technical assistance building vacuum and optical components, and T. Rvachov of the Massachusetts Institute of Technology for his assistance with the ``Cicero Word Generator'' control software.

%===============================================================================================
%\section*{Appendix}
\appendix
\setcounter{section}{0} % reset subsection counter

%------------------------------------------------------------------------------------------------
\section{Ellipse fitting methods}
\label{app:Ellipse}

In this appendix, we give some background regarding ellipse-fitting techniques and illustrate the problem of parameter bias for two different fitting algorithms.

The general form of an ellipse in a cartesian plane is described by the algebraic equation for a conic
\be
  \label{F(a,x)}
  F(\bm{\lambda},\bm{y}) = \bm{\lambda} \cdot \bm{y} = \mathcal{A} y_1^2 + \mathcal{B} y_1 y_2 + \mathcal{C} y_2^2 + \mathcal{D} y_1 + \mathcal{E} y_2 + \mathcal{F} = 0,
\ee
provided that $\mathcal{B}^2 < 4\mathcal{AC}$. Here, $\bm{\lambda} = \{\mathcal{A,B,C,D,E,F}\}$ and $\bm{y} = \{y_1^2, y_1 y_2, y_2^2, y_1, y_2, 1\}$. The center, orientation, major and minor axes of the ellipse are determined by the elements of $\bm{\lambda}$, and the differential phase can be shown to be
\be
  \label{phid2}
  \phi_d = \cos^{-1} \left( -\frac{\mathcal{B}}{2 \sqrt{\mathcal{A} \mathcal{C}}} \right).
\ee
Generally, two types of ellipse-fitting algorithms exist: those that seek to minimize (i) an algebraic distance or (ii) a geometric / orthogonal distance between the ellipse and the data points. While algebraic methods tend to be simple, efficient and can guarantee an ellipse solution to the conic equation \eqref{F(a,x)} (\ie parabolic and hyperbolic solutions can be eliminated), they tend to suffer highly from bias in the ellipse parameters---resulting in a poor fit under certain circumstances. Geometric methods are usually much more accurate than algebraic algorithms, but at the cost of more complexity, more computation and less stability. Since minimizing the orthogonal distance between a point and an ellipse has no closed-form solution, these routines resort to iterative techniques that are not guaranteed to converge on an ellipse.

A commonly used algebraic method is the simple and robust ``direct ellipse fitting'' (DEF) method developed by Fitzgibbon \etal \cite{Fitzgibbon1999} that minimizes the sum of squared algebraic distances between the points and the ellipse, $\sum_{i = 1}^N F(\bm{\lambda},\bm{y}_i)^2$, subject to the constraint $\mathcal{B}^2 = 4\mathcal{AC} - 1$. Recently, Szpak \etal \cite{Szpak2012, Szpak2014} developed an algorithm based on the optimization of the approximate maximum likelihood distance which seeks a balance between the costly geometric methods and stable algebraic techniques. This algorithm---termed the ``fast guaranteed ellipse fitting'' (FGEF) method---also includes error estimation for the geometrically meaningful ellipse parameters (center coordinates, axes and orientation) which we have extended to include an estimate of the differential phase error, $\delta \phi_d$.

%--------------------------------------------
\begin{figure}[!tb]
  \centering
  \subfigure{\includegraphics[width=0.35\textwidth]{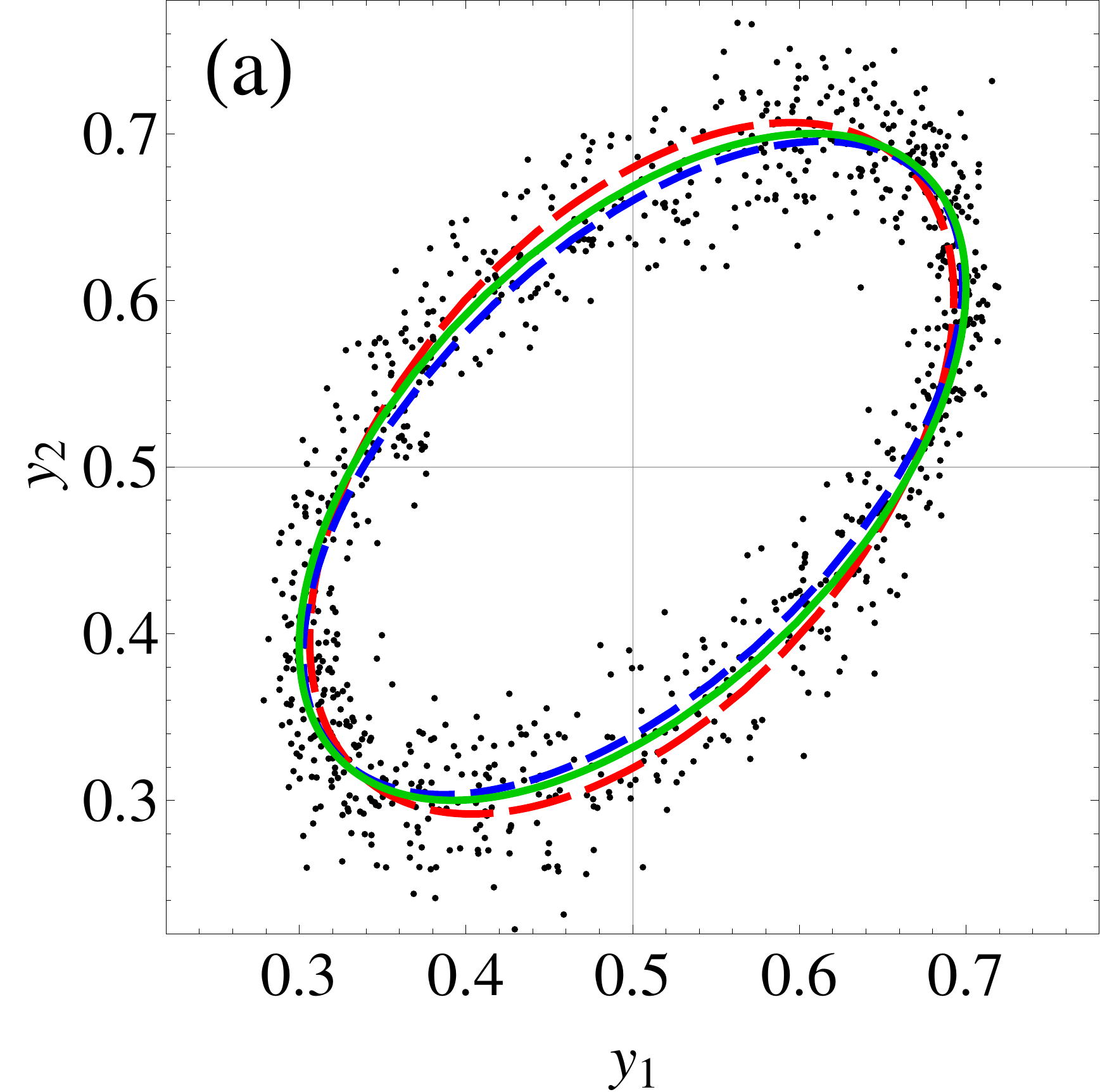}}
  \hspace{0.2cm}
  \subfigure{\includegraphics[width=0.36\textwidth]{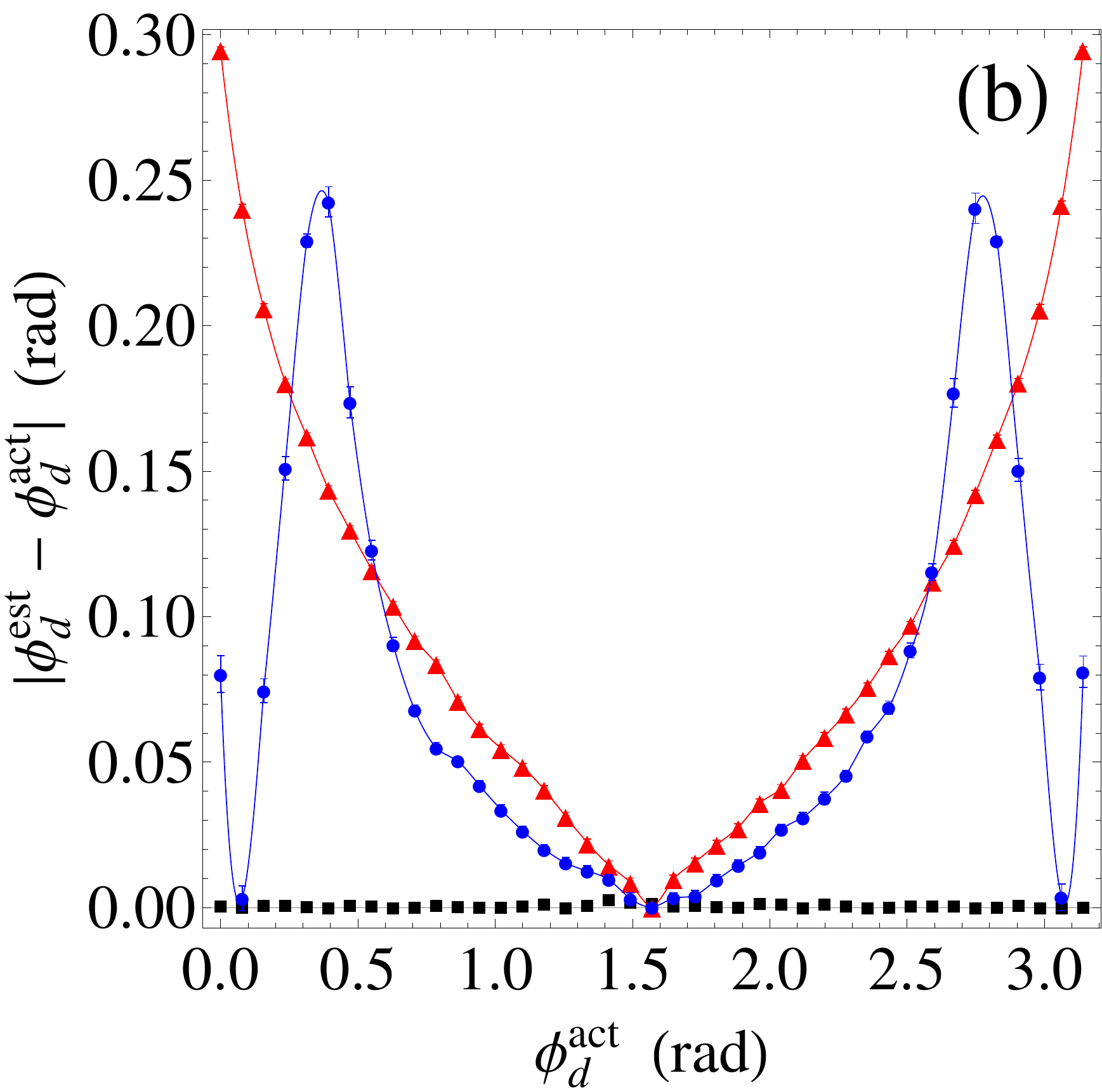}}
  \caption{(a) Synthetic data following an ellipse with added offset noise. The solid green curve represents the actual ellipse, and fits to the data using the DEF method (red curve with big dashes) and FGEF method (blue curve with small dashes). The simulated ellipse contains 500 points with Gaussian-distributed noise on the offset parameters $B_j$ with standard deviations $\{\sigma_{B_1},\sigma_{B_2}\} = \{0.01,0.03\}$ (corresponding to SNR $\sim \{20, 6\}$). Ellipse parameters: $A_1 = A_2 = 0.2$, $B_1 = B_2 = 0.5$, $\kappa = 1$, $\phi_d = 1$ rad. (b) Measured bias in differential phase estimates, $\phi_d^{\rm est}$, from the DEF (red triangles) and FGEF (blue points) methods relative to the actual value, $\phi_d^{\rm act}$. The black squares show the estimates from the differential FRAC method for comparison. On all plots, the error bars correspond to the statistical distribution of fits to 100 synthetic data sets.}
  \label{fig:EllipseFits+OffsetBias}
\end{figure}
%--------------------------------------------

Figure \ref{fig:EllipseFits+OffsetBias} illustrates the bias introduced on the differential phase estimated by the DEF and FGEF methods. For moderate amounts of noise in the offset, the DEF method tends to produce fits that are characteristically compressed along the major axis and stretched along the minor axis of the ellipse, as shown by the red curve in \Fig \ref{fig:EllipseFits+OffsetBias}(a). This effect results in a biased estimate of $\phi_d$ that increases monotonically away from $\pi/2$, as shown in \Fig \ref{fig:EllipseFits+OffsetBias}(b). In contrast to the DEF method, the FGEF algorithm predicts an ellipse (shown in blue) that is much more representative of the actual ellipse (shown in green), and also results in less bias in $\phi_d$ in the central region around $\pi/2$. Outside of this region, the bias behaves nonlinearly in a manner that depends on the ellipse parameters and the level of noise. Here, we point out that these bias estimates are dependent on the type of noise (offset, amplitude, or differential phase) and the amount of noise present in the data, but typically the bias is smallest in the vicinity of $\phi_d = \pi/2$, and decreases with the noise level. In general, ellipse-fitting techniques always generate a non-zero systematic on the differential phase estimate, and depending on the level of sensitivity, this bias must be carefully accounted for when performing precise measurements with $\phi_d$ \cite{Rosi2014, McGuirk2002, Wu2009, Rosi2015, Lamporesi2008}.

%------------------------------------------------------------------------------------------------
\section{Bayesian analysis of Lissajous curves}
\label{app:Bayesian}

In this appendix, we describe in detail our generalized Bayesian analysis technique to estimate the differential phase from Lissajous curves. We also demonstrate the effectiveness of this method using numerically simulated data with Gaussian noise in the offset parameters $\{B_1, B_2\}$ and the differential phase, $\phi_d$. Noise in the amplitude parameters $\{A_1,A_2\}$ of the coupled-sensor model \eqref{yj(a)} can also be included via a trivial modification of the noise model. In what follows, we first provide some relevant theoretical background of the Bayesian estimation technique. For a more comprehensive description of Bayesian analysis in this context, see \Ref \cite{Stockton2007}.

In a generalized system, where $M$ represents a measurement of the system quantities and $V$ represents a variable we are interested in measuring, Bayes' rule can be summarized by the following equation
\be
  \label{BayesRule}
  P(V|M) = \frac{p(V) L(M|V)}{N(M)}.
\ee
Here, $P(V|M)$ is called the ``posterior'' probability distribution and represents our state of knowledge after a measurement, $M$. $p(V)$ is the ``prior'' probability before the measurement, and $L(M|V)$ is called the ``likelihood'' to obtain a certain result for $M$ \emph{given} $V$. The key to the entire estimation process is the likelihood distribution, which is computed based on a specific model of the noise present in the system. The quantity $N(M) = \sum_V L(M|V) p(V)$ is the probability of measuring $M$ integrated over all possible values of $V$, and is just a normalizing factor for the posterior distribution. Mathematically, $L(M|V)$ can be thought of as a function of $V$ with $M$ fixed, and vice versa for $P(V|M)$. The essence of Bayes' rule is that knowledge of the variable $V$ can be updated on a measurement-by-measurement basis---with each successive measurement contributing additional information that narrows the width of the probability distribution associated with $V$. A well-known example of this type of recursive analysis is a Kalman filter \cite{Kalman1960}, which is used extensively in the fields of guidance, navigation and trajectory optimization.

For the specific case of two coupled atom interferometers, the variable of interest is $\phi_d$ and the $i^{\rm th}$ system measurement is given by the pair of (normalized) atomic state populations $M_i = \{n_1, n_2\}_i$. Thus, for a single measurement \Eq \eqref{BayesRule} becomes
\be
  \label{BayesRule2}
  P\big( \phi_d | \{n_1,n_2\}_i \big)_i = \frac{p(\phi_d)_i L\big( \{n_1,n_2\}_i | \phi_d \big)_i}{N\big( \{n_1,n_2\}_i \big)_i},
\ee
where $P(\phi_d | \{n_1,n_2\}_i)_i$ is referred to as the conditional distribution based on the $i^{\rm th}$ measurement. The basic algorithm for Bayes' estimation can be summarized as follows:
\begin{itemize}
  \item[(1)] Choose a suitable initial prior distribution, $p(\phi_d)_{i = 1}$. In our case, we take this to be a uniform distribution within the range $\phi_d \in [0,\pi]$, and zero elsewhere.
  \item[(2)] Record a new measurement $\{n_1,n_2\}_i$, and calculate the likelihood distribution $L(\{n_1,n_2\}_i|\phi_d)_i$ from the noise model.
  \item[(3)] Compute the conditional probability distribution $P(\phi_d|\{n_1,n_2\}_i)_i$ from Bayes' rule \eqref{BayesRule2}.
  \item[(4)] Set the new prior distribution equal to the previous conditional distribution: $p(\phi_d)_{i+1} = P(\phi_d|\{n_1,n_2\}_i)_i$.
  \item[(5)] Repeat steps (2) through (4) until the width of the conditional distribution reduces to the desired level.
  \item[(6)] Estimate the variable of interest, $\phi_d$, using the maximum likelihood value of the final conditional (\ie posterior) probability distribution.
\end{itemize}
This algorithm is illustrated in \Fig \ref{fig:BayesianEstimationExample} in \Sec \ref{sec:Bayesian}.

%------------------------------------------------------------------------------------------------
\subsection*{The likelihood distribution}

\begin{figure}[!tb]
  \centering
  \subfigure{\includegraphics[width=0.35\textwidth]{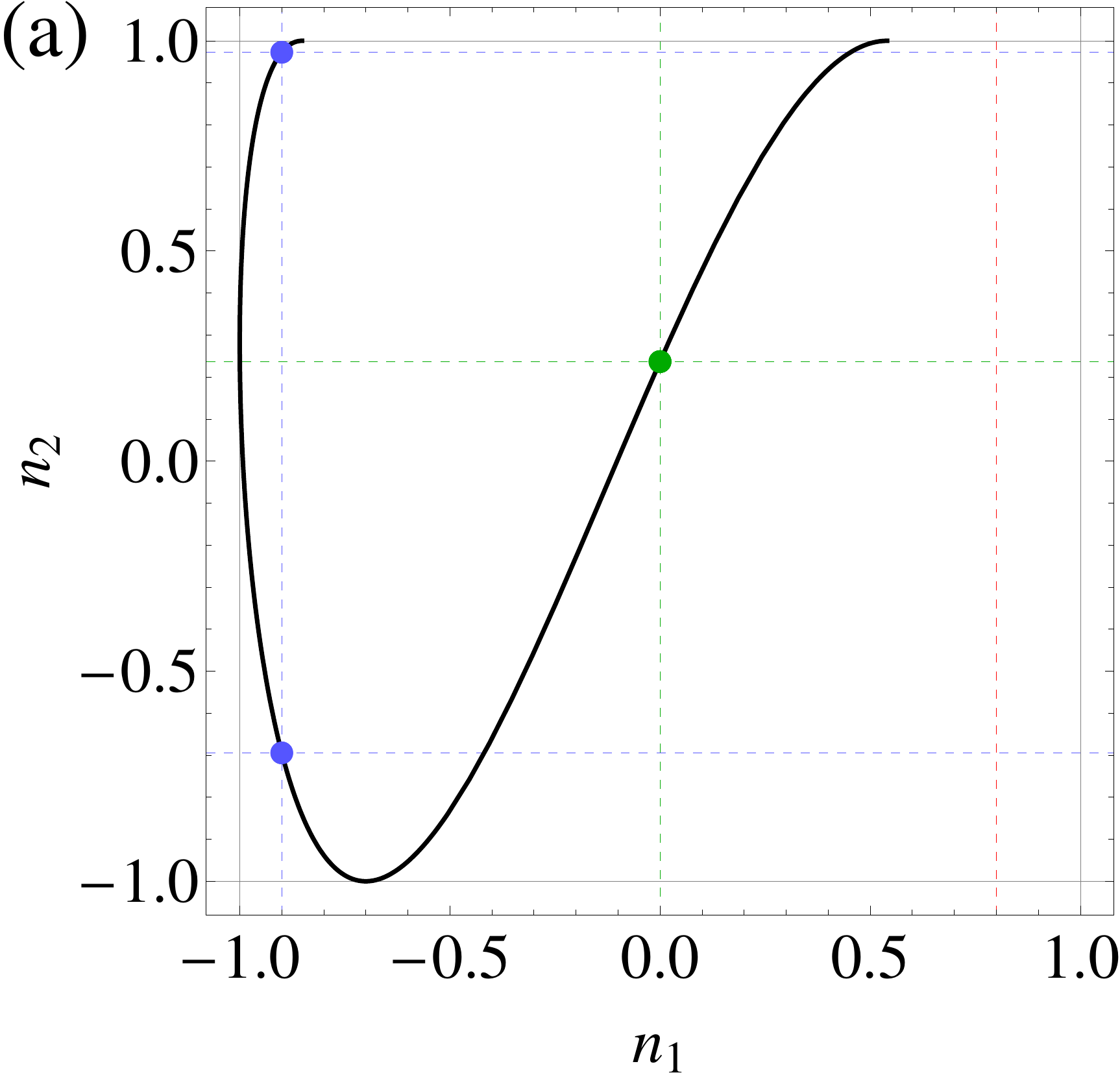}}
  \hspace{0.2cm}
  \subfigure{\includegraphics[width=0.35\textwidth]{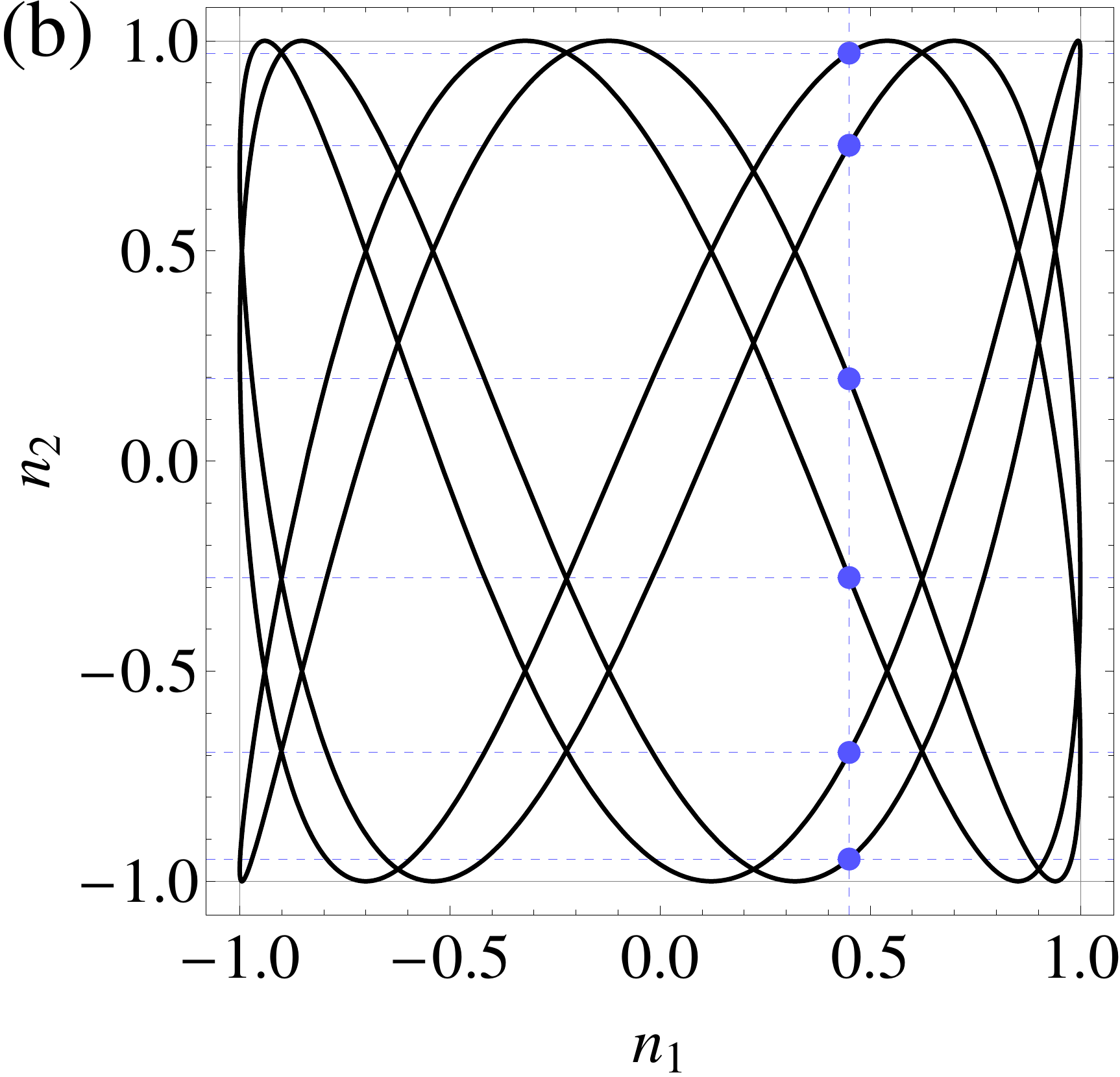}}
  \caption{Example of a Lissajous curve for $\kappa = p/q = 3/7$. (a) The curve is plotted over the phase range $\phi_c \in [0, 2\pi]$ and the solutions $\{n_1, n_2 = s_{2,\ell}\}$ are shown as points. For $n_1 = -0.9$ there are two solutions $s_{2,\ell}$, shown as the blue points. Similarly, for $n_1 = 0$ there is only one possible value of $n_2$ (shown in green), and for $n_1 = 0.8$ no solutions exist. (b) The same Lissajous curve plotted over $\phi_c \in [0, 2\pi q]$ with $q = 7$. In this range, there are always 6 solutions $s_{2,\ell}$ for each value of $n_1$ (although they may not be unique).}
  \label{fig:LissajousSolutions}
\end{figure}

The main challenge in Bayesian analysis is to compute the likelihood distribution $L\big( \{n_1,n_2\} | \phi_d \big)$ given a specific model for $n_1$ and $n_2$. For the specific case of coupled interferometers, there are three possible sources of noise: amplitude, offset and differential phase. To illustrate each source, we modify the definitions of the $n_j$ in \Eq \eqref{nj(phic)} to explicitly include these noise terms
\begin{subequations}
\label{nj(phic)-Noise}
\begin{align}
  n_1(\phi_c) & = (1 + \delta A_1) \cos(\kappa \phi_c + \phi_d + \delta \phi_d) + \delta B_1, \\
  n_2(\phi_c) & = (1 + \delta A_2) \cos(\phi_c) + \delta B_2.
\end{align}
\end{subequations}
The parameters $\delta A_j$, $\delta B_j$, and $\delta \phi_d$ represent uncorrelated noise in the amplitude, offset and differential phase, respectively, each of which is assumed to follow a Gaussian probability distribution with zero mean and non-zero standard deviation. Using this model, the likelihood distribution can be shown to be \cite{Stockton2007}
\be
  \label{Likelihood1}
  L\big( \{n_1,n_2\} | \phi_d \big)
  = \sum_{\ell} \int_{-1}^{1} \frac{P( n_1 | s_1 ) P( n_2 | \{s_{2,\ell}; s_1, \phi_d \})}{\sqrt{1-s_1^2}} \diff s_1.
\ee
Here, $P(n_1|s_1)$ and $P( n_2 | \{s_{2,\ell}; s_1, \phi_d \})$ are the single-sensor conditional probability distributions for $n_1$ and $n_2$, which we discuss in more detail below. The quantities $s_1 \equiv \cos(\kappa \phi_c + \phi_d)$ and $s_{2,\ell} \equiv \cos(\phi_c)$ are the principle variables on which the coupled measurements $n_1$ and $n_2$ depend in the model \eqref{nj(phic)-Noise}. Due to the periodic nature of the Lissajous equations \eqref{nj(phic)}, for each value of $n_1$ there are multiple possible solutions for $n_2$ (as shown in \Fig \ref{fig:LissajousSolutions}). We assign an integer $\ell$ to each of these solutions. More specifically, $s_{2,\ell}$ is the $\ell^{\rm th}$ root of $n_2$ \emph{given} $n_1 = s_1$. The sum over $\ell$ appearing in \Eq \eqref{Likelihood1} accounts for all possible solutions. In the distribution functions $P(n_1|s_1)$ and $P( n_2 | \{s_{2,\ell}; s_1, \phi_d \})$, we denote the implicit dependence on variables $s_1$ and $\phi_d$ by a semi-colon. This notation emphasizes that the quantity $s_{2,\ell}$ is coupled to $s_1$ through the common phase $\phi_c$ \footnote{Since it is assumed that $\phi_c$ is random and unknown, the probability distributions of $s_1$ and $s_{2,\ell}$ are equivalent to that of a sinusoid: $P(s_{2,\ell}|\phi_c) = (1 - s_{2,\ell}^2)^{-1/2}$.}. Finally, we point out that the coupled variables $s_1$ and $s_{2,\ell}$ both depend on $\phi_d$, but we do not write this dependence explicitly.

At this point, we need to know the possible values $n_2 = s_{2,\ell}$ (given a measurement of $n_1 = s_1$) which enter into the likelihood distribution. We devote the remainder of this section to a detailed description of computing the roots of the Lissajous equations \eqref{nj(phic)}. As mentioned above, due to the non-linear nature of Lissajous curves, there are multiple possible solutions for $n_2$ given a single value of $n_1$ within a predefined phase range. We denote these solutions $s_{2,\ell}$ for integer $\ell$. When $\kappa = 1$, the Lissajous curve collapses to an ellipse, and only two values of $n_2$ exist for each $n_1$ over any $2\pi$ range of $\phi_c$. In this case, it is straightforward to compute the two solutions as $s_{2,\pm 1} = \cos[\cos^{-1}(s_1) \pm \phi_d]$. However, when $\kappa \not= 1$, the problem is much more complex. If the scale factor ratio can be written in the form $\kappa = p/q$, where $p$ and $q$ are prime numbers, then the period of the Lissajous curve is $2\pi q$---requiring $q$ revolutions to form a closed loop. Within each $2\pi$ interval, there can be either 0, 1 or 2 solutions of $n_2$ for each $n_1$, as illustrated in \Fig \ref{fig:LissajousSolutions}.

%--------------------------------------------
%--------------------------------------------

To calculate these solutions for a given $n_1 = s_1$ and $\phi_d$, it is necessary to know the approximate range of common phase spanned by the data: $\phi_c \in [\phi_c^{\rm min}, \phi_c^{\rm max}]$ \footnote{In practice, this range estimate does need not to be very precise---we find that estimating the correct range to within $\pm \pi$ still results in a precise estimate for $\phi_d$. However, overestimating the phase range may result in a slower convergence rate for the estimate. See \Sec \ref{sec:KRbCorrelation} for a description of how the phase range can be estimated experimentally.}. With this information, we compute the range of phase spanned by sensor 1, $\theta \in \kappa [\phi_c^{\rm min}, \phi_c^{\rm max}] + \phi_d$, and we subdivide this range into intervals of $\pi$ such that the $\ell^{\rm th}$ interval is defined as the range $\theta_{\ell} \in [\ell, \ell+1) \pi$, where $\ell = \lfloor \theta/\pi \rfloor$. Here, the brackets $\lfloor \cdots \rfloor$ indicate the floor function. Beginning with the left-most interval, we check for solutions sequentially at each $\pi$ phase bin until the entire range is spanned. Empirically, we find that if a solution exists within the $\ell^{\rm th}$ interval given a value $n_1$, then it is unique and can be written explicitly as
\be
  \label{n2,l}
  s_{2,\ell} = \left\{
  \begin{array}{ll}
    \cos[(\cos^{-1}s_1 - \phi_d + 2\pi m_{1,\ell})/\kappa] & \mbox{ for even } \ell, \\
    \cos[(\cos^{-1}s_1 + \phi_d - 2\pi m_{2,\ell})/\kappa] & \mbox{ for odd } \ell,
  \end{array} \right.
\ee
where the integers $m_{1,\ell}$ and $m_{2,\ell}$ are defined as
\be
  m_{1,\ell} = \left\{
  \begin{array}{cc}
	\lfloor (\ell - 1)/2 \rfloor & \ell < -1, \\
    0                            & -1 \le \ell < 2, \\
    \lfloor \ell/2 \rfloor       & \ell \ge 2,
  \end{array} \right.
  \;\;\;\;\;
  m_{2,\ell} = \left\{
  \begin{array}{cc}
	\lfloor \ell/2 \rfloor       & \ell < -2, \\
    0                            & -2 \le \ell < 1, \\
	\lfloor (\ell + 1)/2 \rfloor & \ell \ge 1.
  \end{array} \right.
\ee
With these solutions in hand, it is possible to compute the likelihood \eqref{Likelihood1} given specific noise models for the single-sensor probability distributions $P(n_1|s_1)$ and $P(n_2|\{s_{2,\ell};s_1,\phi_d\})$. We now investigate the specific cases of offset and differential phase noise on the extraction of $\phi_d$ from simulated data sets. This analysis can also be extended to include noise in the fringe amplitudes through the parameters $\delta A_j$ \cite{Stockton2007}, but we do not consider this case here.

%------------------------------------------------------------------------------------------------
\subsection*{Offset noise}

%--------------------------------------------
\begin{figure*}[!tb]
  \centering
  \includegraphics[width=0.96\textwidth]{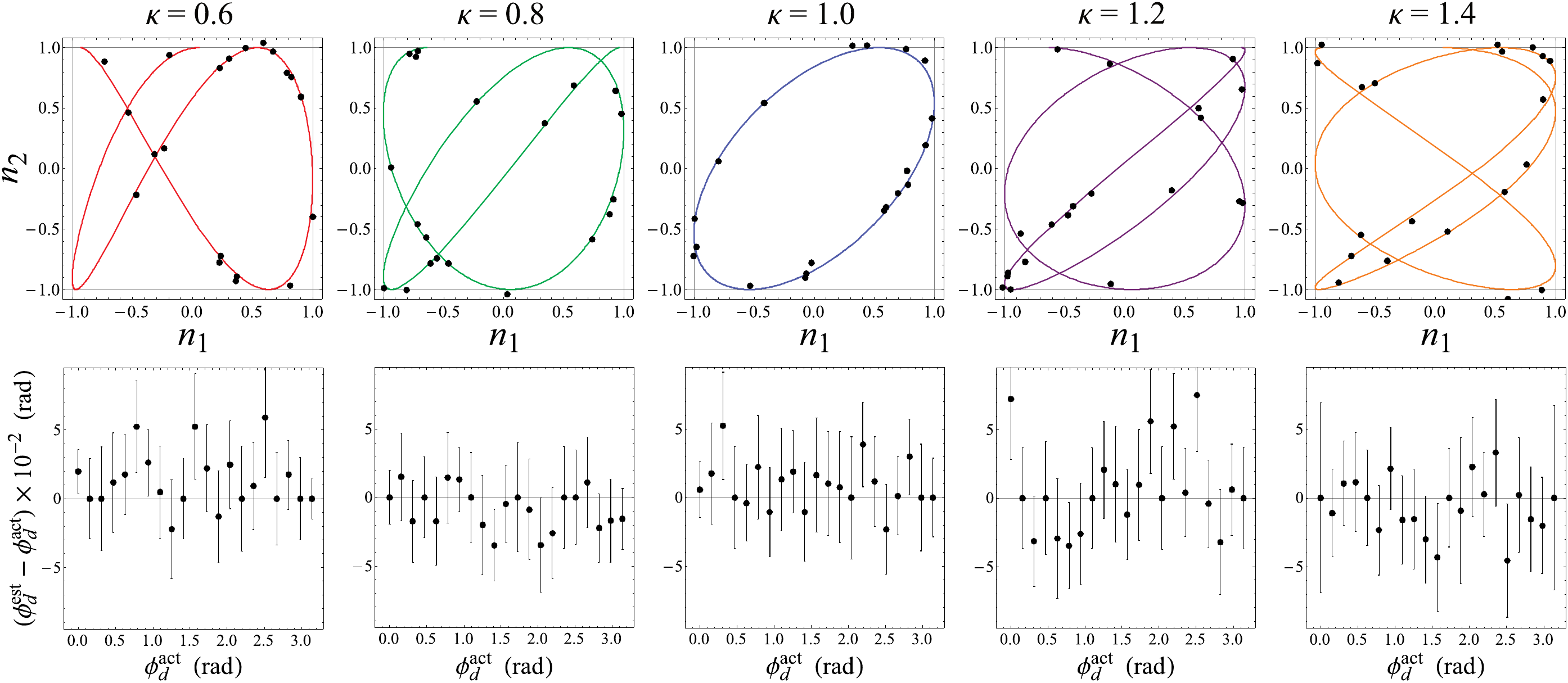}
  \caption{Simulated Bayesian phase estimates in the presence of offset noise. Top row: simulated data for $\kappa = \{0.6, 0.8, 1.0, 1.2, 1.4\}$. Here, points based on \Eq \eqref{nj(phic)-Noise} were generated from common phases chosen randomly over the range $\phi_c \in [-2\pi, 2\pi]$. The differential phase was set to $\phi^{\rm act}_d = 1$ rad, and offset noise was applied to the points via Gaussian distributions with standard deviations $\{\sigma_{B_1},\sigma_{B_2}\} = \{0.02,0.04\}$---corresponding to SNR $\sim 50$ and 25, respectively. Bottom row: the systematic error in the Bayesian phase estimate, $\phi_d^{\rm est} - \phi_d^{\rm act}$, as a function of $\phi_d^{\rm act}$ for each value of $\kappa$. Only 10 points were required to bring the statistical uncertainty indicated by the error bars to $\lesssim 50$ mrad.}
  \label{fig:BayesianErrorVsAlpha-OffsetNoise}
\end{figure*}
%--------------------------------------------

When the system exhibits noise only in the offset of the atomic state measurements, the parameters $\delta B_j$ are randomly distributed for each repetition of the experiment, and $\delta A_j = 0$ and $\delta \phi_d = 0$ in the model \eqref{nj(phic)-Noise}. Under realistic conditions, these noise parameters follow a Gaussian distribution with zero mean and standard deviations given by $\sigma_{B_j}$, and the single-sensor conditional probabilities can be written as
\begin{subequations}
\begin{align}
  P(n_1|s_1) & \propto \exp\big[-(n_1 - s_1)^2/2\sigma_{B_1}^2\big], \\
  P(n_2|\{s_{2,\ell}; s_1, \phi_d\}) & \propto \exp\big[-(n_2 - s_{2,\ell})^2/2\sigma_{B_2}^2\big].
\end{align}
\end{subequations}

Figure \ref{fig:BayesianErrorVsAlpha-OffsetNoise} shows some examples of simulated data in the presence of offset noise, where the differential phase has been extracted using the Bayesian estimation algorithm described above. These simulations show that $\phi_d$ can be precisely estimated over the full range of $0 - \pi$, and for a wide variety of scale factor ratios. Here, we demonstrate the technique for the limited range $\kappa \in [0.6,1.4]$, but we have also verified that the extraction method works well outside this range. In contrast to ellipse-fitting techniques, no systematic bias in the phase estimates is observed, and fewer points are required to converge to competitive error levels.

%------------------------------------------------------------------------------------------------
\subsection*{Differential phase noise}

Since the noise parameter associated with the differential phase, $\delta \phi_d$, adds directly to the quantity of interest, $\phi_d$, we can account for this type of noise by adding an extra convolution with our noise model at the end of any likelihood calculation. We choose to examine the case of Gaussian noise for the differential phase, such that the conditional probability distribution is
\be
  P(\phi_d'|\phi_d) \propto \exp[-(\phi_d' - \phi_d)^2/2\sigma_{\phi_d}^2].
\ee
Here, $\phi_d'$ represents a measured value of the differential phase in the presence of Gaussian noise centered on the most likely value, $\phi_d$, and $\sigma_{\phi_d}$ is the standard deviation of the noise distribution. The modified likelihood function is described by the convolution
\be
  \label{Likelihood-Convolution}
  L(\{n_1,n_2\}|\phi_d) \propto \int_{-\infty}^{\infty} L(\{n_1,n_2\}|\phi_d') P(\phi_d'|\phi_d) \diff \phi_d'.
\ee
%
%--------------------------------------------
\begin{figure*}[!tb]
  \centering
  \includegraphics[width=0.96\textwidth]{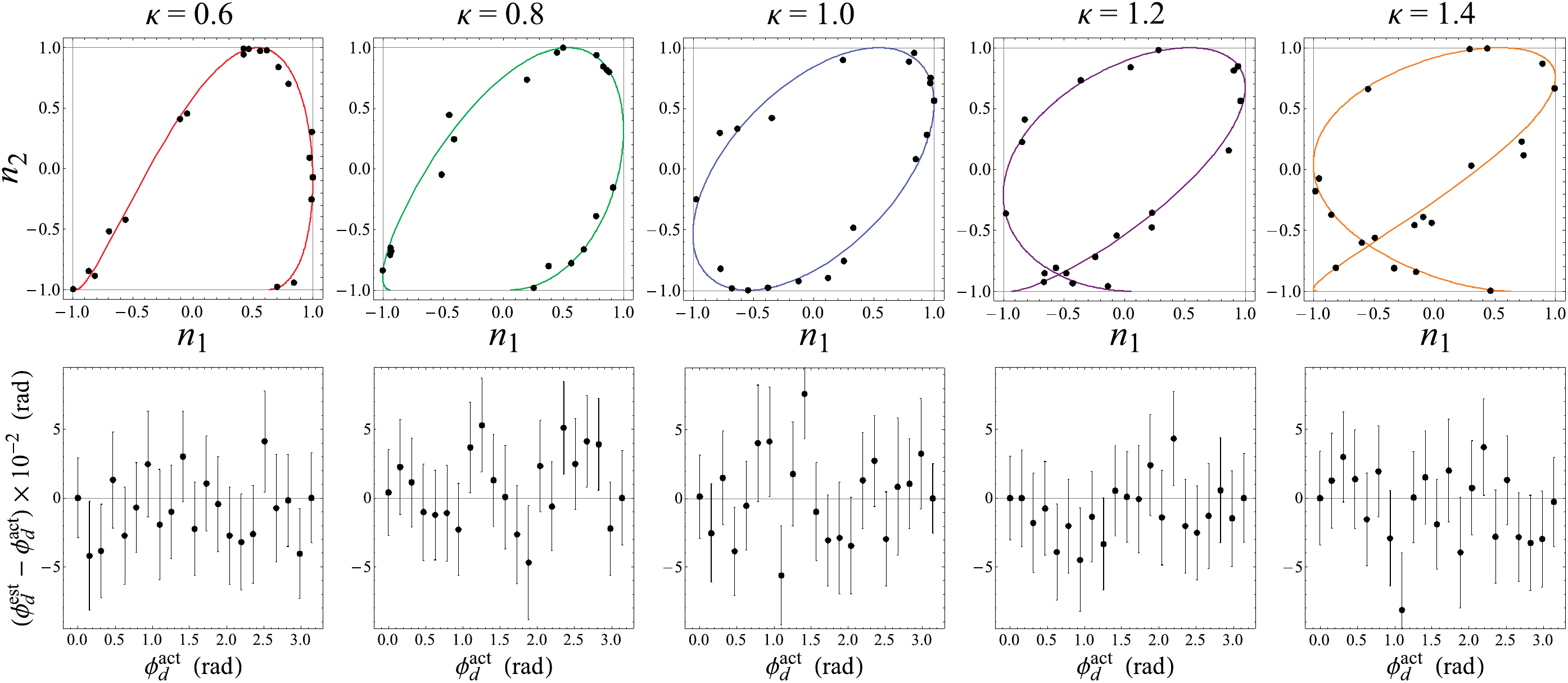}
  \caption{Simulated Bayesian phase estimates in the presence of differential phase noise. Top row: simulated data for $\kappa = \{0.6, 0.8, 1.0, 1.2, 1.4\}$. Similar to \Fig \ref{fig:BayesianErrorVsAlpha-OffsetNoise}, common phases are randomly chosen over the range $\phi_c \in [-\pi, \pi]$, and the differential phase is set to $\phi^{\rm act}_d = 1$ rad. Differential phase noise is applied to each point with a standard deviation of $\sigma_{\phi_d} = 0.1$ rad (SNR $\sim 10$). Bottom row: the systematic error in the Bayesian phase estimate, $\phi_d^{\rm est} - \phi_d^{\rm act}$, as a function of $\phi_d^{\rm act}$ for each value of $\kappa$. Only 10 points were required to bring the statistical uncertainty to $\lesssim 50$ mrad.}
  \label{fig:BayesianErrorVsAlpha-DiffPhaseNoise}
\end{figure*}
%--------------------------------------------
%
In a similar fashion to the offset, in the absence of any other noise sources it is necessary to estimate multiple candidate solutions for $\phi_d$ over a given range of $\phi_c$ in order to compute the likelihood function. Before convolving with the conditional probability distribution in \Eq \eqref{Likelihood-Convolution}, the likelihood function can be written as
\be
  L(\{n_1,n_2\}|\phi_d') = \sum_{k} \delta(\phi_d' - \phi_{d, k}),
\ee
where $\delta(x)$ is the Dirac delta function, and the sum over $k$ accounts for all candidate solutions $\phi_{d,k}$ that exist in the common phase range $\phi_c \in [\phi_c^{\rm min}, \phi_c^{\rm max}]$. These solutions can be computed by, again, dividing the phase range into intervals of $\pi$, and labeling each of them by an integer $k = \lfloor \phi_c/\pi \rfloor$. We find that two possible solutions exist for $\phi_d$ within each interval, which we denote as $\phi_{d,k}^{(\pm)}$ for $\phi_c \in [\pm k \pi, \pm(k + 1) \pi)$. Explicitly, these phases can be computed from
\be
  \phi_{d,k}^{(\pm)} = \cos^{-1}(n_1) \pm \kappa[\cos^{-1}(n_2) - 2\pi m_{k}],
\ee
where $m_{k} = (-1)^{k} \lfloor (|k| + 1)/2 \rfloor$. We transform these phases into the range of $0 - \pi$ using $\phi_{d,k}^{(\pm)} \to \cos^{-1}[\cos(\phi_{d,k}^{(\pm)})]$. Finally, this result is convolved with the Gaussian noise model to obtain
\be
  L(\{n_1,n_2\}|\phi_d) = \sum_{k} \exp[-(\phi_d - \phi_{d,k}^{(\pm)})^2/2\sigma_{\phi_d}^2].
\ee

Two subtleties exist with this analysis, however, that warrant discussion. First, when the common phase range exceeds $\phi_c \in [-\pi, \pi]$, the Bayesian analysis may predict multiple equally probable values for $\phi_d$. This is obviously a problem if we are interested in a precise, unique estimate of the differential phase, and we have no pre-existing knowledge of its value. Therefore, we restrict our consideration of the problem to a range of common phase within $-\pi$ to $\pi$. Second, the noise parameter $\delta \phi_d$ can theoretically take any value, \ie $\delta \phi_d \in (-\infty, \infty)$, although in practice it is limited to a finite range defined by $\sigma_{\phi_d}$. So far, we have considered $\phi_d$ only in the range of 0 to $\pi$, but for situations where $\sigma_{\phi_d} \gtrsim \pi/4$, the likelihood distribution can have significant contributions from the wings of the adjacent $\pi$ phase intervals. This effect can be taken into account by using the fact that $P(\phi_d) = P(-\phi_d) = P(2\pi - \phi_d)$, and adding mirrored versions of the likelihood to the convolution in \Eq \eqref{Likelihood-Convolution}. This ``tiling'' technique can be extended to account for large noise levels, where more than one $\pi$ phase bin is spanned \cite{Stockton2007}.

Figure \ref{fig:BayesianErrorVsAlpha-DiffPhaseNoise} shows some examples of simulated data in the presence of differential phase noise. As for the case of offset noise, estimates of $\phi_d$ exhibit no significant bias over the full range of $0-\pi$, and for a large range of scale factor ratios. Additionally, only a small number of points are required to converge to a level of uncertainty less than that of the noise defined by $\sigma_{\phi_d}$. The convergence of this uncertainty as a function of the number of measurements is the subject of the next section.

%------------------------------------------------------------------------------------------------
\subsection*{Scaling with measurement number}

\begin{figure}[!tb]
  \centering
  \subfigure{\includegraphics[width=0.35\textwidth]{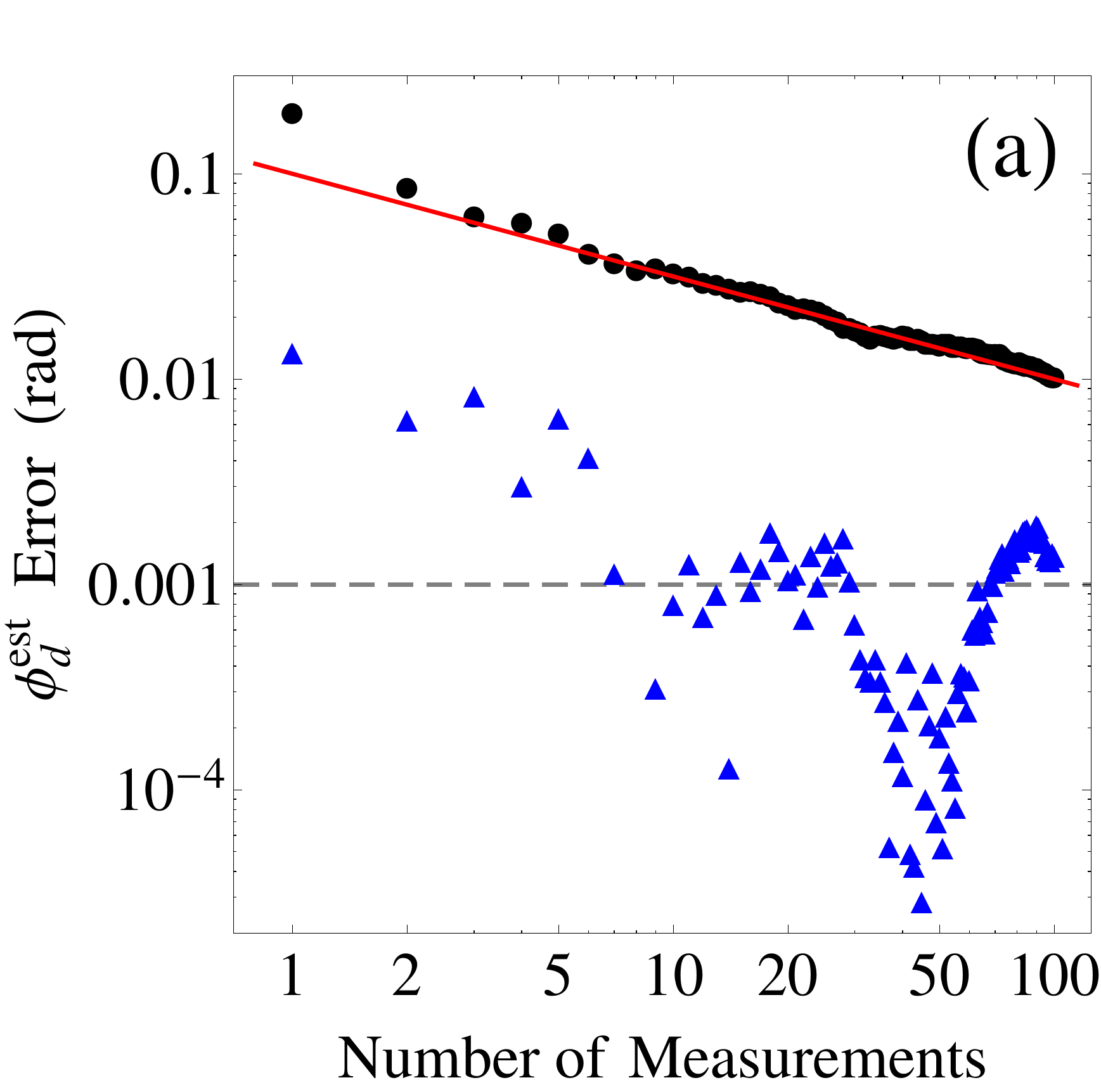}}
  \hspace{0.2cm}
  \subfigure{\includegraphics[width=0.35\textwidth]{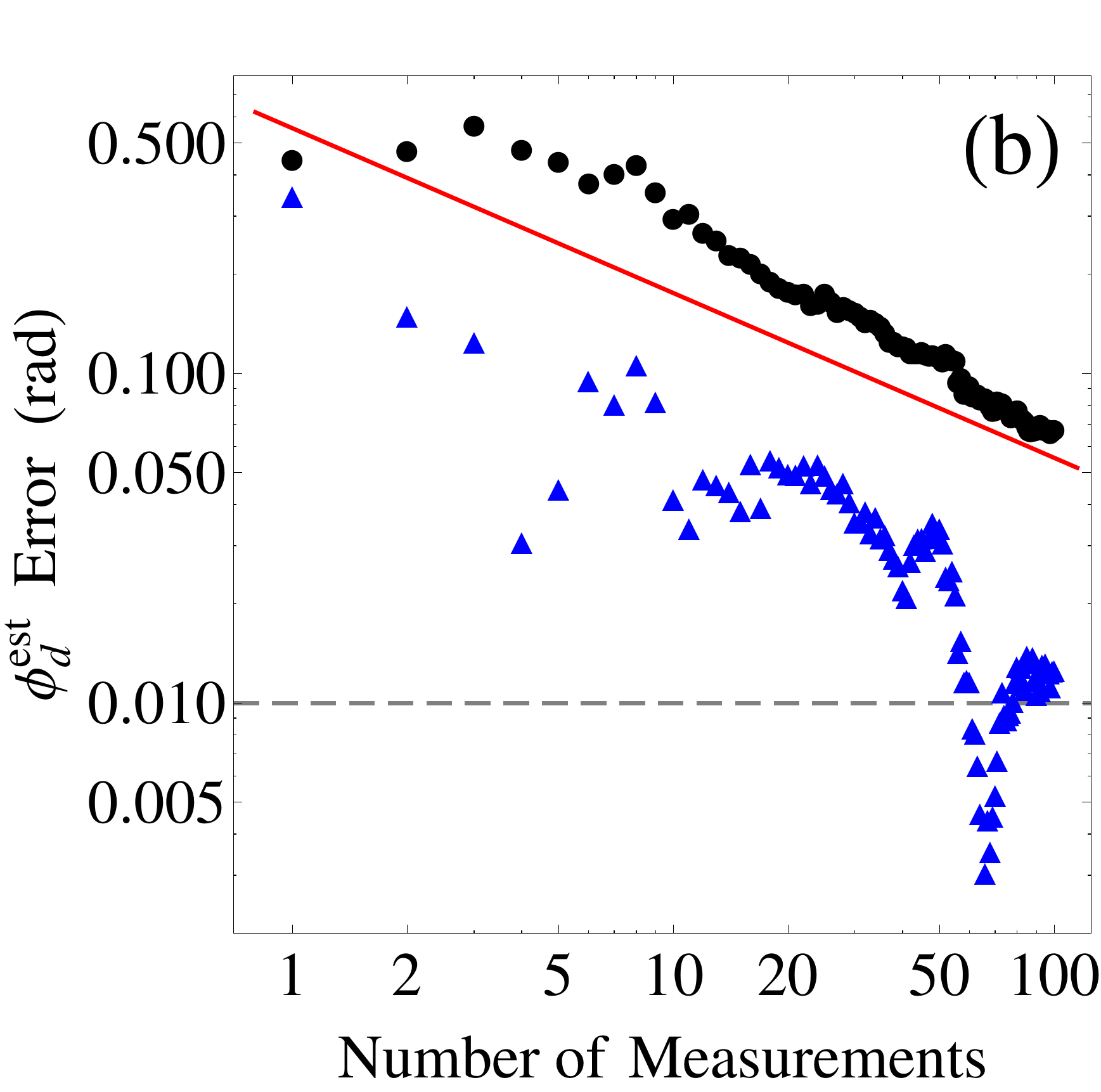}}
  \caption{The statistical and systematic error in the Bayesian phase estimator as a function of the number of measurements, $N$, used in the Bayesian analysis for (a) differential phase noise and (b) offset noise. In both plots, $\kappa = 0.8$, $\phi_d^{\rm act} = 1$ rad, and $M = 50$ samples were used. Black points represent the statistical uncertainty $\epsilon^{\rm stat}_{\phi_d}$, and blue triangles indicate the systematic error $\epsilon^{\rm sys}_{\phi_d}$. The solid red lines indicate the minimum convergence rates based on $\sigma_{\phi_d}/\sqrt{N}$ for differential phase noise and on $1/\sqrt{N\,I(\phi_d)}$ for offset noise, where $I(\phi_d)$ is the Fisher information given by \Eq \eqref{I(phid)}. The dashed horizontal lines represent the nominal phase resolution used in the simulations. Values of $\sigma_{\phi_d} = 0.1$ rad (SNR $\sim 10$) and $\sigma_{B_1} = \sigma_{B_2} = 0.2$ (SNR $\sim 5$) were used as the noise parameters in (a) and (b), respectively.}
  \label{fig:Convergence}
\end{figure}

To test the scaling of the statistical and systematic error of the Bayesian estimator as a function of the number of measurements, we performed the following study. We randomly generated $M = 50$ samples of ``measurements'', each containing 100 points following the model \eqref{nj(phic)-Noise} with noise added to either the differential phase or the offset. As a function of the measurement number, $N$, within each sample, we computed the Bayesian estimate $\phi_d^{\rm est}(N)$ and the standard deviation of the associated probability distribution $\delta \phi_d^{\rm est}(N)$. The statistical error for each measurement is taken as the average of $\delta \phi_d^{\rm est}(N)$ over all $M$ samples, which we denote as $\epsilon_{\phi_d}^{\rm stat}(N) = \langle \delta \phi_d^{\rm est}(N) \rangle_M$. Similarly, the systematic error is defined as $\epsilon_{\phi_d}^{\rm sys}(N) = \langle |\phi_d^{\rm est}(N) - \phi_d^{\rm act}| \rangle_M$. The results are shown in \Fig \ref{fig:Convergence}.

For the specific case of noise that contributes directly to the variable of interest (\eg differential phase noise) the statistical uncertainty of the Bayesian estimator is given by $\epsilon_{\phi_d}^{\rm stat} = \sigma_{\phi_d}/\sqrt{N}$. As we show in \Fig \ref{fig:Convergence}(a), the measured statistical error closely follows this dependence. Similarly, on average the systematic error drops to a level much less than $\epsilon_{\phi_d}^{\rm stat}$ after only a few measurements. This level is primarily determined by the grid resolution used when computing the likelihood distribution for $\phi_d$. During the estimation procedure, we initially set the phase grid resolution to $\sim \pi/100$, and we refine this grid size on a measurement-by-measurement basis. As the likelihood distribution narrows, grid points are redistributed toward the maximum likelihood value. We find that this grid optimization procedure can improve the resolution by up to an order of magnitude (depending on the level of noise in the system), while keeping the number of integral evaluations per measurement fixed.

%--------------------------------------------
%--------------------------------------------

For the more general case of noise present in a parameter that is indirectly related to the quantity of interest through some function, the uncertainty is constrained by the Cramer-Rao lower bound
\be
  \epsilon_{\phi_d}^{\rm stat} \geq \frac{1}{\sqrt{N \, I(\phi_d)}}.
\ee
This relationship can be used to compute the minimum convergence rate of the statistical error in the presence of offset or amplitude noise, for example, where the noise affects $\phi_d$ indirectly through the quantities $\{n_1,n_2\}$. The Cramer-Rao lower bound includes the Fisher information, $I(\phi_d)$, of an individual measurement, which can be computed from the likelihood distribution $L(\{n_1,n_2\}|\phi_d)$ as follows
\be
  \label{I(phid)}
  I(\phi_d)
  = - \left< \frac{\partial^2}{\partial \phi_d^2} \ln \big[ L(\{n_1,n_2\}|\phi_d) \big] \right>_{\{n_1,n_2\}}
  = \iint \frac{\diff n_1 \diff n_2}{L(\{n_1,n_2\}|\phi_d)} \left( \frac{\partial L}{\partial \phi_d} \right)^2.
\ee
Here, the brackets $\langle \cdots \rangle_{\{n_1,n_2\}}$ denote an average over the random variables $\{n_1,n_2\}$. The Fisher information is a measure of the amount of information that a random variable (or a set of random variables) carries about an unknown parameter. In this case, the unknown parameter of interest is $\phi_d$ and the set of random variables is the set of measurements $\{n_1,n_2\}$, which are governed by the likelihood distribution $L(\{n_1,n_2\}|\phi_d)$---hence its appearance in \Eq \eqref{I(phid)}. This quantity has no closed-form expression for the case of offset or amplitude noise in our system, and must be evaluated numerically. For the parameters used in \Fig \ref{fig:Convergence}(b), we find $I(\phi_d) \simeq 3.3$, which gives a minimum convergence rate of $0.55/\sqrt{N}$. This rate is consistent with the measured statistical uncertainties shown in the figure. We note that the Fisher information empirically scales as $I \sim e^{-\beta \sigma_B}$, where $\beta$ is a large factor that depends on the differential phase and the scale factor ratio used (\eg $\beta \sim 35$ for $\kappa = 0.8$ and $\phi_d = 1$ rad). Thus, with only a moderate reduction to the level of offset noise in the system, one can dramatically improve the convergence rate of the Bayesian estimate.

%------------------------------------------------------------------------------------------------
\section{Response of a dual-species interferometer to mirror vibrations}
\label{app:Response}

Here, we summarize the essential theoretical tools required to evaluate the response of both single- and dual-species interferometers to vibrational noise of the retro-reflection mirror.

First, we provide a review of the sensitivity function for a single atom interferometer, $g(t)$. This function characterizes how the interferometer transition probability behaves in the presence of fluctuations in the Raman laser phase difference, $\varphi_{\rm L}(t)$. Developed previously for use with atomic clocks \cite{Dick1987}, the sensitivity function is a useful tool that can be applied, for example, to evaluate the response of the interferometer to laser phase noise \cite{Cheinet2008}, or to correct for spurious vibrations in the Raman beam optics \cite{Barrett2014, Geiger2011phd, Geiger2011}. We are primarily interested in the latter.

The sensitivity function is a unitless quantity that is defined as follows
\be
  \label{g(t)}
  g(t)
  = \lim_{\delta\varphi \to 0} \frac{\delta \Phi(\delta \varphi, t)}{\delta\varphi}
  = 2 \lim_{\delta\varphi \to 0} \frac{\delta P(\delta\varphi, t)}{\delta\varphi},
\ee
where $\delta \varphi$ is a phase jump occurring at time $t$ during the interferometer that modifies the total interferometer phase, $\Phi$, by an amount $\delta\Phi$, and the transition probability $P(\Phi) = (1 - \cos \Phi)/2$ by a corresponding amount $\delta P$. Thus, the interferometer phase due to an arbitrary phase noise function, $\varphi(t)$, can be computed as
\be
  \Phi_{\varphi}
  = \int_{-\infty}^{\infty} g(t) \diff \varphi(t)
  = \int_{-\infty}^{\infty} g(t) \frac{\diff \varphi(t)}{\diff t} \diff t.
\ee
The quantum mechanical nature of the atom plays a crucial role on the sensitivity function---in particular, the evolution of the internal atomic states during each Raman pulse. Using the procedure outlined in \Refs \cite{Barrett2014, Cheinet2008}, the sensitivity function, $g_j(t)$, of an interferometer with timing parameters labeled with subscript ``$j$'' can be shown to be
\be
  \label{gj(t)}
  g_j(t) = \left\{
  \begin{array}{cc}
    -\sin\big(\Omega_j^{\rm eff} (t - \Delta T_j)\big)        & 0 < t - \Delta T_j \le \tau_j, \\
    -1                                              & \tau_j < t - \Delta T_j \le T_j + \tau_j, \\
    -\sin\big(\Omega_j^{\rm eff}(t - \Delta T_j - T_j)\big)   & T_j + \tau_j < t - \Delta T_j \le T_j + 3\tau_j, \\
    1                                               & T_j + 3\tau_j < t - \Delta T_j \le 2T_j + 3\tau_j, \\
    -\sin\big(\Omega_j^{\rm eff}(t - \Delta T_j - 2T_j)\big)  & 2T_j + 3\tau_j < t - \Delta T_j \le 2T_j + 4\tau_j, \\
    0                                               & \mbox{otherwise}.
  \end{array}
  \right.
\ee
Here, $T_j$ is the interrogation time, $\tau_j$ is a pulse duration, $\Omega_j^{\rm eff}$ is the effective Rabi frequency associated with the two-photon Raman transitions, and $\Delta T_j$ is a delay with respect to $t = 0$ that facilitates a difference in the start time between interferometers. It is assumed that $\Omega_j^{\rm eff} \tau_j = \pi/2$, such that the first and third interferometer pulses have pulse areas of $\pi/2$ with duration $\tau_j$, and the second is a $\pi$-pulse of duration $2\tau_j$.

To evaluate the response of an interferometer to Raman mirror motion, the phase noise function is first expressed as $\varphi_j(t) = k_j^{\rm eff} z(t)$, with $z(t)$ representing the time-dependent position of the mirror along the axis of the beams. Then, the phase shift of interferometer $j$ due to movement of the Raman mirror is
\be
  \label{phijvib-2}
  \phi_j^{\rm vib}
  = \int_{-\infty}^{\infty} w_j(t) a^{\rm vib}(t) \diff t
  = k_j^{\rm eff} \int_{-\infty}^{\infty} f_j(t) a^{\rm vib}(t) \diff t,
\ee
where $a^{\rm vib}(t) = \ddot{z}(t)$ is the time-dependent acceleration of the mirror due to vibrations, $w_j(t) = k_j^{\rm eff} f_j(t)$ is a time-dependent weight function for the mirror accelerations, and $f_j(t)$ is called the response function associated with the $j^{\rm th}$ interferometer. This function is given by the integral of the sensitivity function: $f_j(t) = -\int_{0}^{t} g_j(t') \diff t'$, and can be evaluated as
\be
  \label{fj(t)}
  f_j(t) = \left\{
  \begin{array}{cc}
    \frac{1}{\Omega_j^{\rm eff}} \big( 1 - \cos\Omega_j^{\rm eff} (t - \Delta T_j)  \big) &
    0 < t - \Delta T_j \le \tau_j, \\

    (t - \Delta T_j) + \frac{1}{\Omega_j^{\rm eff}} - \tau_j &
    \tau_j < t - \Delta T_j \le T + \tau_j, \\

    T_j + \frac{1}{\Omega_j^{\rm eff}} \big(1 - \cos \Omega_j^{\rm eff}(t - \Delta T_j - T) \big) &
    T_j + \tau_j < t - \Delta T_j \le T_j + 3\tau_j, \\

    2T_j + 3\tau_j + \frac{1}{\Omega_j^{\rm eff}} - (t - \Delta T_j) &
    T_j + 3\tau_j < t - \Delta T_j \le 2T_j + 3\tau_j, \\

    \frac{1}{\Omega_j^{\rm eff}} \big( 1 - \cos \Omega_j^{\rm eff}(t - \Delta T_j - 2T_j) \big) &
    2T_j + 3\tau_j < t - \Delta T_j \le 2T_j + 4\tau_j, \\

    0
    & \mbox{otherwise}.
  \end{array}
  \right.
\ee
%
%--------------------------------------
\begin{figure}[!tb]
  \centering
  \includegraphics[width=0.60\textwidth]{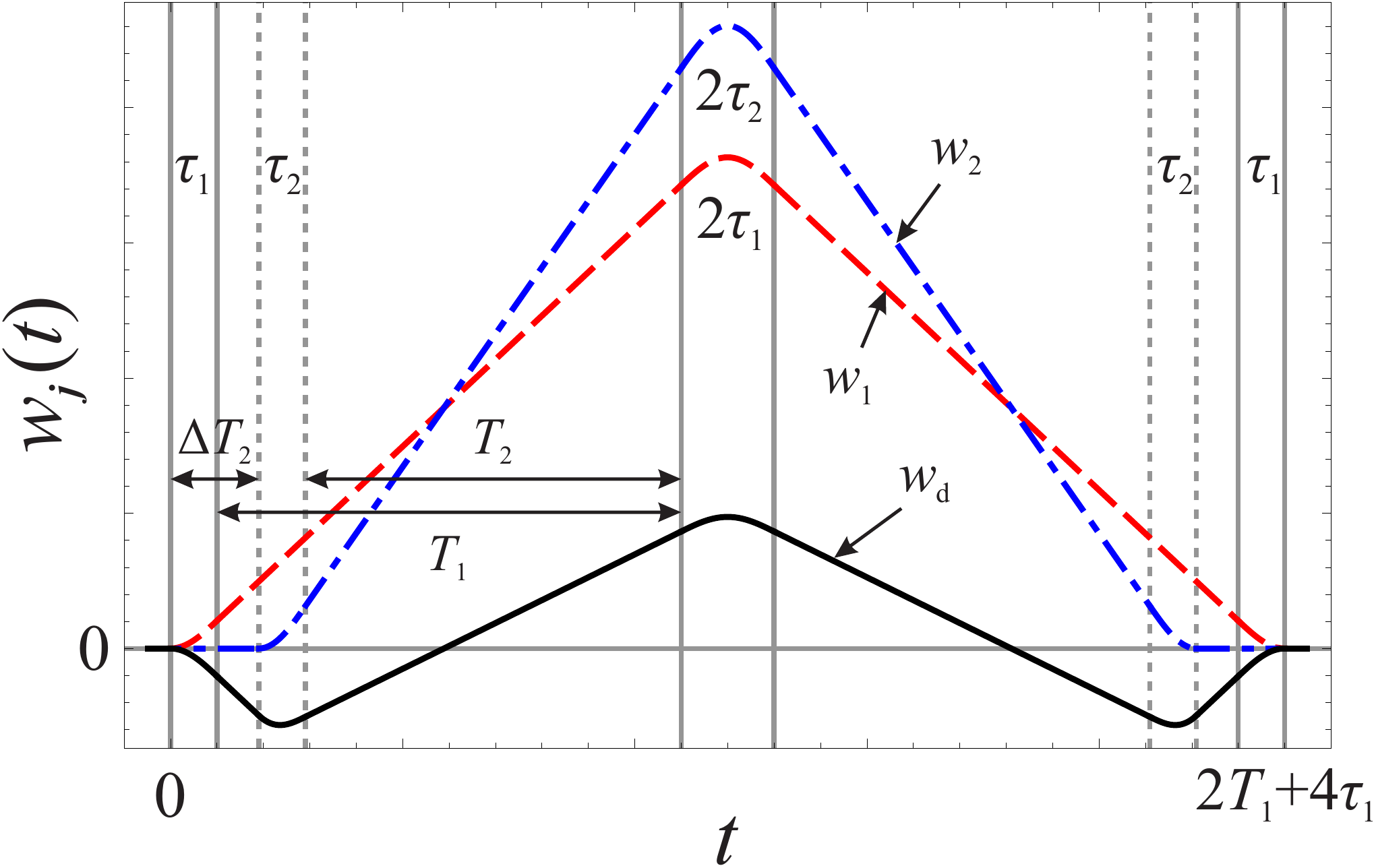}
  \caption{Weight functions, $w_j(t)$, described by the response function \eqref{fj(t)}. These weights determine the phase shift associated with mirror vibrations in \Eq \eqref{phijvib-2}. The pulse durations, $\tau_j$, satisfy $\Omega_j^{\rm eff} \tau_j = \pi/2$. The differential weight function, \ie the difference between the red and blue curves, is shown in black.}
  \label{fig:wj(t)}
\end{figure}
%--------------------------------------
%
At its heart, \Eq \eqref{phijvib-2} is a generalization of the well-known interferometer phase shift due to a constant acceleration, $a$ %
\be
  \phi_j
  = S_j a
  = k_j^{\rm eff} (T_j + 2\tau_j) \left( T_j + \frac{4\tau_j}{\pi} \right) a
  \simeq k_j^{\rm eff} T_j^2 \, a.
\ee
In this relation, the quantity $S_j \simeq k_j^{\rm eff} T_j^2$ is equivalent to the integral of the weight function, $w_j(t)$, which determines how strongly the mirror vibration at time $t$ contributes to the interferometer phase shift. This function is triangle-shaped, as shown in \Fig \ref{fig:wj(t)}, which indicates that the phase contributions are smallest near $t = \Delta T_j$ and $\Delta T_j + 2T_j + 4\tau_j$, where the wavepacket separation is a minimum. Similarly, the weights are largest near the mid-point, $t = \Delta T_j + T_j + 2\tau_j$, where the separation between the interfering states is a maximum.

For the case of two coupled interferometers, the differential phase shift resulting from mirror vibrations can be expressed as
\be
  \label{phidvib}
  \phi_d^{\rm vib} = \phi_1^{\rm vib} - \phi_2^{\rm vib} = \int_{-\infty}^{\infty} w_d(t) a^{\rm vib}(t) \diff t,
\ee
where the differential weight function, $w_d(t)$, is given by the difference between the single-sensor weight functions
\be
  w_d(t) = w_1(t) - w_2(t) = k_1^{\rm eff} f_1(t) - k_2^{\rm eff} f_2(t).
\ee
This function has an intuitive understanding. For the extreme case when $k_1^{\rm eff} = k_2^{\rm eff}$ and the two interferometers are perfectly overlapped (\ie $\Delta T_1 = \Delta T_2$, $T_1 = T_2$, $\tau_1 = \tau_2$), $w_d(t)$ is zero everywhere. This implies that the differential phase shift due to mirror motion is $\phi_d^{\rm vib} = 0$---corresponding to perfect common-mode phase noise rejection. In the opposite extreme, when either $k_1^{\rm eff} \not= k_2^{\rm eff}$ or the interferometers are not well-overlapped, vibration noise induces a differential phase shift $\phi_d^{\rm vib}$ between the two sensors given by \Eq \eqref{phidvib}. This non-zero phase shift is directly responsible for uncorrelated contributions to $\phi_d$ in the case of non-overlapped interferometers, and it explains the loss of common-mode rejection in the case of coupled interferometer with different scale factors. For the case of a constant acceleration, \Eq \eqref{phidvib} can also be used to derive the systematic shift $\delta \phi_d^{\rm sys} = (S_1 - S_2) a$ resulting from interferometers exhibiting $S_1 \not= S_2$.

\begin{figure}[!tb]
  \centering
  \includegraphics[width=0.96\textwidth]{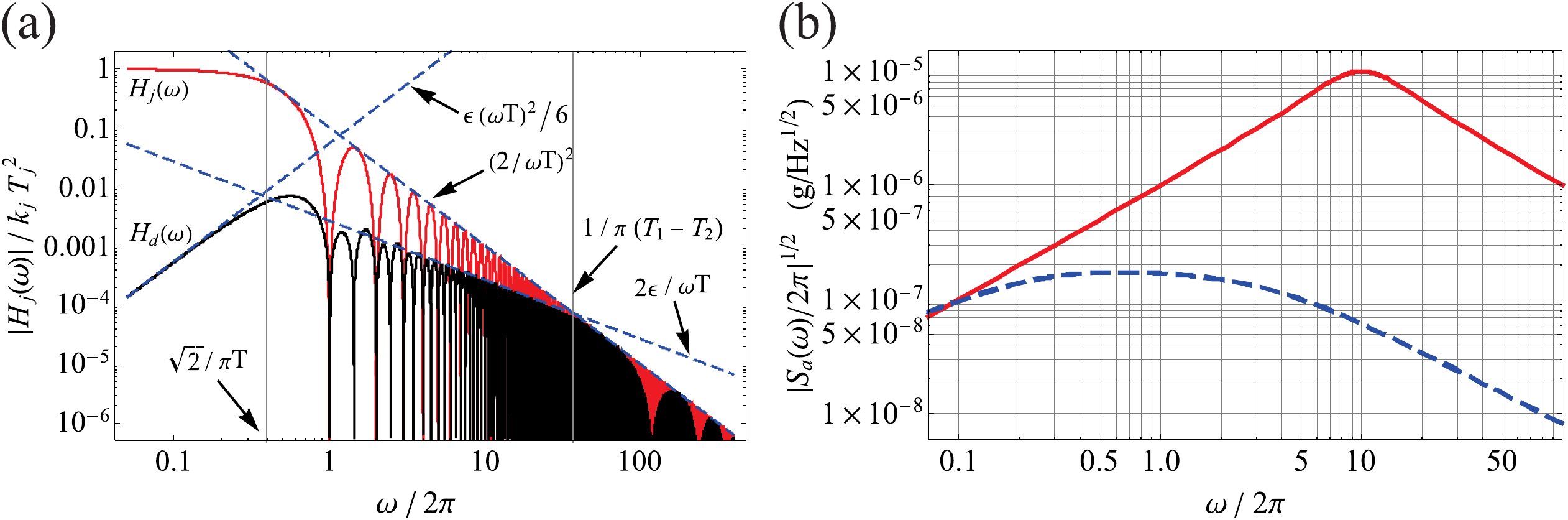}
  \caption{(a) Normalized transfer functions, $|H_j(\omega)|/k_j^{\rm eff} T_j^2$, described by \Eq \eqref{Hj(omega)} for coupled K-Rb interferometers. These functions determine the response of the single-species (red curve) and dual-species (black curve) interferometers to acceleration noise at different frequencies, $\omega$. Here, $T_1 = 1$ s and $T_2 = (1-\epsilon) T_1$, with $\epsilon = 1 - \sqrt{k_1^{\rm eff}/k_2^{\rm eff}} \simeq 0.0087$, and $\tau_1 = \tau_2 = 10$ $\mu$s. (b) Model curves for the power spectral density of ground vibrations described by \Eq \eqref{Sa(omega)}. The solid red curve corresponds to the power spectral density of a ``quiet'' location \cite{LeGouet2008, Merlet2009} that we simulated with the parameters $\xi = 0.5$, $\chi = 1$, $\omega_0/2\pi = 10$ Hz, $a_{\rm rms} = 1.4 \times 10^{-4}\,g$. The dashed blue curve corresponds to ``low-noise'' conditions achievable with passive vibration isolation ($\xi = 2$, $\chi = 0.8$, $\omega_0/2\pi = 1$ Hz, $a_{\rm rms} = 1.4 \times 10^{-6}\,g$). These two curves were used to compute the rms phase noise in table \ref{tab:SensitivityLimits}.}
  \label{fig:Hj(omega)-Sa(omega)}
\end{figure}

One can characterize how mirror vibrations with a given frequency spectrum affect each interferometer by computing the mean-squared phase noise
\be
  \label{(phi_rms^vib)^2}
  (\phi_{\rm rms}^{\rm vib})^2 = \frac{1}{2\pi} \int_0^{\infty} |H_j(\omega)|^2 \mathcal{S}_a(\omega) \diff \omega.
\ee
Here, $\mathcal{S}_a(\omega)$ is the power spectral density of acceleration noise on the mirror, and $H_j(\omega)$ is the transfer function associated with interferometer $j$ given by the Fourier transform of $w_j(t)$. The transfer function describes how acceleration noise at a given frequency affects the phase over the duration of the interferometer. For frequencies $\omega \ll \Omega_j^{\rm eff}$ and pulse separations $T_j \gg \tau_j$, this function is well-approximated by
\be
  \label{Hj(omega)}
  H_j(\omega) = -i e^{-i\omega(\Delta T_j + T_j + 2\tau_j)} k_j^{\rm eff} T_j^2 \, \mbox{sinc}^2\left( \frac{\omega T_j}{2} \right)
\ee
For the dual-species interferometer, one uses the differential transfer function in the same fashion:
\be
  \label{Hd(omega)}
  H_d(\omega) = H_1(\omega) - H_2(\omega).
\ee
These functions are shown in \Fig \ref{fig:Hj(omega)-Sa(omega)}(a) for realistic experimental parameters associated with a K-Rb interferometer. Here, there is a clear difference between the transfer functions associated with single-species and dual-species interferometers. For the individual sensors, the transfer function is well-approximated by the square of a sinc function---which exhibits regular zeroes at the fundamental frequency $1/T_j$ and an envelope that decreases as $(2/\omega T_j)^2$. This dependence implies that the interferometer naturally filters the high-frequency components of the vibration spectrum, with a $-3$ dB cut-off frequency of $\omega_j^{\rm cut}/2\pi = \sqrt{2}/\pi T_j \simeq 1/2T_j$.

%--------------------------------------
%--------------------------------------

The differential transfer function, on the other hand, has a much more complicated frequency dependence. We will focus on the most interesting case for WEP tests, \ie when the wave vectors satisfy $k_1^{\rm eff} = (1 - \epsilon)^2 k_2^{\rm eff}$ where $\epsilon \ll 1$, and the two interferometers are symmetrically overlapped in time as shown in \Fig \ref{fig:wj(t)}. Under these conditions, we find that $H_d$ can be approximated by
\begin{align}
\begin{split}
  \label{Hd(omega)2}
  |H_d(\omega)|
  & \approx k_1^{\rm eff} (T_1^2 - T_2^2) \mbox{sinc}\left( \frac{\omega(T_1 + T_2)}{2} \right) \mbox{sinc}\left( \frac{\omega(T_1 - T_2)}{2} \right) \\
  & - 2 \epsilon k_1^{\rm eff} T_2^2 \mbox{sinc}^2\left( \frac{\omega T_2}{2} \right).
\end{split}
\end{align}
It follows that there is a competition between the two terms in this expression. For the extreme case when $\epsilon = 0$ (\ie $k_1^{\rm eff} = k_2^{\rm eff}$), the differential transfer function is dominated by the first term, which is identically zero for all frequencies only if $T_1 = T_2$. This represents the ideal case for gravity gradiometry applications. On the other hand, when $\epsilon > 0$ and $T_1 = T_2$, the second term in \Eq \eqref{Hd(omega)2} dominates. Since the two interferometers are assumed to have different wave vectors, it is not possible to make the transfer function zero at all frequencies. However, it is straightforward to show that $H_d = 0$ in DC provided that $k_1^{\rm eff} T_1^2 = k_2^{\rm eff} T_2^2$. This criteria optimizes the rejection of common-mode vibration noise at frequencies below the cut-off for a single-sensor, $\omega_j^{\rm cut}$, and can be achieved by adjusting the interrogation times such that $T_2 = \sqrt{k_1^{\rm eff}/k_2^{\rm eff}} T_1 = (1 - \epsilon) T_1$.

Figure \ref{fig:Hj(omega)-Sa(omega)}(a) shows a comparison between single-sensor and differential transfer functions for $T \sim 1$ s interferometers. When operated differentially, the sensitivity to vibrations at frequencies less than $\omega_j^{\rm cut}$ is typically more than 3 orders of magnitude below that of the single interferometer, despite the fact that $k_1^{\rm eff} \not= k_2^{\rm eff}$.

Figure \ref{fig:Hj(omega)-Sa(omega)}(b) displays the power spectral density function, $S_a(\omega)$, that was used to compute the rms phase noise in table \ref{tab:SensitivityLimits}. These curves are based on a regression model for ground accelerations \cite{Matsuda2008} described by
\be
  \label{Sa(omega)}
  \mathcal{S}_a(\omega)
  = N(\xi,\chi) \left( \frac{4\xi\omega_0\omega^3}{(\omega^2 - \omega_0^2)^2 + (2 \xi \omega_0 \omega)^2} \right)^\chi
  \, \frac{2\pi a_{\rm rms}^2}{\omega}.
\ee
This model has a single peak at $\omega = \omega_0$ and contains two positive shape parameters, $\xi$ and $\chi$, that determine the sharpness of the peak and the scaling of the wings of the distribution, respectively. The quantity $a_{\rm rms}^2$ is the mean-squared acceleration of the corresponding time-domain acceleration signal, $a(t)$, and $N(\xi, \chi)$ is a normalization factor that depends on the shape parameters. This factor is chosen such that the integral $\int_{-\infty}^{\infty} \mathcal{S}_a(\omega) \diff \omega = 2\pi a_{\rm rms}^2$.

%================================================================================================
\bibliographystyle{iopart-num}
\bibliography{ICEBib}
%%%%%%%%%%%%%%%%%%%%%%%%%%%%%%%%%%%%%%%%%%%%%%%%%%%%%%%%%%%%%%%%%%%%%%%%%%%%%%%%%%%%%%%%%%%%%%%%%
\end{document}